\documentclass{aa}
\usepackage{epsfig}
\usepackage{natbib}
\usepackage{amssymb}
\bibpunct{(}{)}{;}{a}{}{,}

\newcommand{\bez}{\begin{eqnarray*}}
\newcommand{\eez}{\end{eqnarray*}}
\newcommand{\be}{\begin{equation}}
\newcommand{\ee}{\end{equation}}
\newcommand{\beq}{\begin{eqnarray}}
\newcommand{\eeq}{\end{eqnarray}}
\newcommand{\bc}{\begin{center}}
\newcommand{\ec}{\end{center}}

\def \Delnud{\Delta \nu_{\rm D}}

\def\mpr{m_{\rm p}}

\def \rmd{{\rm d}}
\def \pc{p^{\rm c}}
\def \fw{f_{\rm H_2O}}

\def \fd{f_{\rm d}}
\def \fdh{f_{\rm d}^{\rm h}}
\def \fdc{f_{\rm d}^{\rm c}}
\def \alw{\alpha_{\rm L}}
\def \water{H$_2$O}
\def \ald{\alpha_{\rm d}}
\def \aldh{\alpha_{\rm d}^{\rm h}}
\def \aldc{\alpha_{\rm d}^{\rm c}}

\def \Nd{N_{\rm d}}
\def \Nh{N_{\rm H_2}}
\def \Nw{N_{\rm H_2O}}

\def \Tst{T_{\rm *}}
\def \Kabs{K_{\rm abs}}
\def \Kabsave{\overline{K}_{\rm abs}}
\def \Kabsh{K_{\rm abs}^{\rm h}}
\def \Kabsc{K_{\rm abs}^{\rm c}}
\def \Kabsnu{K_{{\rm abs}, \lambda}}

\def \rhod{\rho_{\rm d}}

\def \Bef{\tilde{B}}
\def \beef{\tilde{\beta}}

\def \Td{T_{\rm d}}
\def \Tdh{T_{\rm d}^{\rm h}}
\def \Tdc{T_{\rm d}^{\rm c}}

\def \Ad{A_{\rm d}}
\def \mum{\mu{\rm m}}

\def \tauc{\tau_{\rm c}}
\def \taum{\tau_{\rm m}}

\def \cminv3{{\rm cm}^{-3}}

\def\up{{u}}
\def\low{{l}}
\def\Cij{C_{ul}}
\def\Cji{C_{lu}}
\def\Aij{A_{ul}}
\def\Aji{A_{lu}}
\def\pji{p_{lu}}
\def\pij{p_{ul}}
\def\pdij{p^{\rm c}_{ul}}
\def\pdji{p^{\rm c}_{lu}}

\def\lambdaij{\lambda_{ul}}
\def\nuij{\nu_{ul}}
\def\nij{n_{ul}}
\def\nji{n_{lu}}
\def\ni{n_{\up}}
\def\nj{n_{\low}}
\def\gi{g_{\up}}
\def\gj{g_{\low}}

\begin{document}

\title{A self-consistent  model of a 22 GHz water maser in a dusty environment
near late-type stars}

\author{Natalia Babkovskaia and Juri Poutanen}
\institute{Astronomy Division, P.O.Box 3000, FIN-90014 University of Oulu, Finland \\
 \email{natalia.babkovskaia@oulu.fi; juri.poutanen@oulu.fi} }

\titlerunning{Water masers in late-type stars}
\authorrunning{Babkovskaia \& Poutanen}

%\email{nbabkovs@sun3.oulu.fi}

%\date{Received ; accepted }
\date{\today}

\abstract{ We study the conditions for operation of the 22 GHz
ortho-water maser in a dusty medium near late-type stars. The main
physical processes, such as exchange of energy between dust and
gas in the radiation field of a star, radiative cooling by water
molecules and pumping of water masers are described
self-consistently. We show that the presence of dust grains of
various types (or of one type with size distribution) strongly
affects the maser action. The pumping mechanism based on the
presence of the dust of different optical properties is able to
explain water masers in the silicate carbon star V778 Cyg.
However, the masers in  the winds from asymptotic giant branch
stars require an additional source of heating, for instance due to
the dust drift through the gas.
  }

\maketitle
%------------------------------------------------------------------

\section{Introduction}

22 GHz ($6_{16} \rightarrow 5_{23}$) ortho-water maser emission
has been detected from many astronomical objects, such as
late-type stars, star formation regions, and active galactic
nuclei. In recent years, much progress in studying maser objects
has been made due to increasing angular resolution and sensitivity
of radio-interferometers, which allow one to obtain detailed maps
of the distribution of maser sources. These observations  provide
important information about the structure, physical conditions and
kinematics of the gas in these sources. The  observational data
show that in most of them, maser radiation comes from the region
co-spatial with sources of the infrared dust emission
\citep{DG81,YD00,DB94} and, therefore,  maser amplification may
take place in a dusty environment.

In asymptotic giant branch  stars (AGBs) and red supergiants
(RSGs), masers are mostly observed in stellar winds. Water masers
observed in some silicate carbon stars, e.g. V778 Cyg
\citep{En94,EL94}, are not associated with the winds, but most
probably arise in an oxygen-rich disk around a companion star in a
binary system \citep{SS05}.

The first model of water masers in  late-type stars was proposed
by \citet{De77}, who considered an expanding, spherically
symmetric circumstellar envelope. The line photons were assumed to
escape from the wind due to the high velocity gradients. The gas
temperature was not computed self-consistently, but was assumed to
be a power-law function of distance from the star. In his model,
the strongest water masers are predicted along tangential
directions where the coherent length is largest. He used
approximate formulae for the collisional cross-sections and
neglected the influence of dust on the maser strength.

\citet{CE85} have presented a more elaborate model employing  a
new set of collisional cross-sections by \citet{Gr80} and using
the wind parameters taken from the model of \citet{GS76}. In this
model, the acceleration of the wind is provided by the radiation
pressure on the dust particles, the temperature was assumed to be
determined by the stellar radiation, while the gas is viscously
heated by collisions with faster moving dust grains. \citet{CE85}
assume the same temperature profile for outflows with very
different characteristics. They have accounted for the influence
of dust {\em emission} (using a simplified dust absorption
coefficient) on the excitation of water molecule, but ignored the
effect of line photon absorption by the dust. The escape of
photons was again assumed to be determined by the velocity
gradient.

\citet{HY01} also simulated masers in the winds in a \citet{So60}
approximation, using more recent collisional cross-sections
provided by \citet{GM93}. They obtained, however,  suspiciously
high gas temperatures because of ignoring molecular cooling.  The
influence of dust absorption on the water molecule population was
again neglected.

Recent observations by MERLIN and VLA have shown that the winds
from AGBs  and RSGs are strongly inhomogeneous. Water masers arise
at distances of 10-60 AU from AGBs and have a typical size of
$2-4$ AU across \citep{BC03}, while in RSGs all sizes are 5 to 10
times larger. Moreover, the filling factor of the clouds is only
$\sim 0.01$. Because of the relatively small sizes, the velocity
gradient within the clouds is negligible and the Sobolev
approximation becomes questionable.

The aforementioned maser wind models \citep{De77,CE85,HY01}
operate on the \citet{dJ73} mechanism of collisional pumping where
the breakdown of thermal equilibrium occurs  because the gas
becomes transparent for different transitions at different
physical depths. However, the presence of dust can strongly
influence the maser strength \citep{GK74,Ke75,St77,BS77}.
\cite{GK74} suggested that  the hot dust radiation  can excite
water molecules to the vibrational state, and the heat sink could
be provided by collisions with cooler (than dust) hydrogen
molecules. The possibility of inversion in such a situation was
questioned by \cite{De81}. Alternatively, the cold dust can
produce the necessary inversion \citep[][ hereafter
BP04]{St77,BS77,De81,CK84a,CW95,WW97,YF97,BP04}. In an optically
thick environment, the excitation temperature takes a value
between the dust and the gas temperatures depending on the
relative role of the dust and collisions in the destruction of the
line photons. \cite{De81} has pointed out that water ice can
provide a very effective heat sink due to  a  strong peak near 45
$\mum$ in the absorption coefficient. For other dust types the
inversion of maser level populations is smaller by just  20--30
per cent (BP04).

In the cold dust--hot gas model, an arbitrarily thick layer can
participate in the maser action provided the gas and dust
temperatures differ sufficiently. The temperature difference
between the gas and the dust can appear in late-type stars due to
the viscous gas heating by rapidly moving dust, presence of the
dust particles of different types and sizes which assure their
different temperatures, or shock waves. Thus, the dust present in
the outflow plays a dual role, providing heating of the gas as
well as absorbing line photons and increasing the maser strength.
Because in this situation the photon escape is determined by dust
absorption and the physical size of the maser clump, we can use
the static approximation (\citealt{YF97}; BP04) for maser
modeling. This model  also can be used for masers in a slowly
rotating disk around a companion of the silicate carbon star V778
Cyg
 \citep{SS05}.

The purpose of this paper is to construct a self-consistent model
of water masers in the environment typical for clumpy winds from
AGBs and RSGs or maser sites in the vicinity of silicate carbon
stars. We take into account here the most important heating/
cooling processes for the gas and dust and study  in detail the
influence of various dust types and sizes on the  maser action. In
Section 2, we formulate the problem and give a detailed
description of calculation method. Section 3 is devoted to the
results of calculations and discussion of the multi-temperature
dust pumping mechanism. In section 4, we   illustrate our model,
applying it to the masers near late-type stars. A summary is given
in Section 5.

\section{Formulation of the problem and calculation method}
\label{sec:param}

We investigate the maser effect in medium containing  mixture of
the gas (molecular hydrogen and water vapor) and dust in the black
body  radiation field of a late-type star. We simultaneously solve
the population balance equation for the water molecule and the
radiative transfer equation (RTE) for water spectral lines using
an escape probability method. We also account for the saturation
effect in masing lines. Our approach is similar to that of BP04.

In addition, we  self-consistently compute the dust and gas
temperature accounting for main heating and cooling processes. As
we will see below, the dust temperature is determined mainly by
the radiation field of a star and is independent of the processes
in the gas phase, therefore it can be computed immediately.

The gas temperature, however, depends on the energy exchange with
dust and cooling provided by water vapor, which depends on the
level populations of the water molecule. The populations in turn
depend on the gas (and dust) temperature. Therefore, we implement
the iteration scheme where we first guess the gas temperature,
compute populations and the corresponding cooling rate, then
correct the gas temperature to satisfy the energy balance, until
convergence is reached.

We consider a slab geometry (with half-thickness $H$) which is a
reasonably good representation of clumps in the winds from  AGBs
and RSGs  flattened in the process of formation due to shock waves
and/or due to different expansion velocities in radial and
tangential directions \citep[see arguments in][]{BC03}. It also
can be used to model masers from disks. The slab thickness $2 H$
is varied from about $10^{11}$ cm to $10^{16}$ cm which is the
upper limit corresponding to the total size of the masing region
in RSGs \citep{RM78}. The stellar temperature is taken to be
$\Tst=3000$ K. We consider the hydrogen concentration
$10^{7}\lesssim \Nh \lesssim 10^{12}$ cm$^{-3}$ and {\it mass}
fraction of water $10^{-8} \lesssim \fw=9\Nw/\Nh \lesssim 10^{-2}$
(where $\Nw$ is water concentration and the factor of 9 comes from
the ratio of water to hydrogen masses).

\subsection{Radiative transfer}

The solution of the RTE is obtained by the escape probability
method. The mean intensity of the radiation averaged over the line
profile at the optical depth $\tau$ (measured from the upper
boundary of the slab) can be written in the form (see e.g. BP04)
\beq \label{eq:J_new}
&&\overline{J} (\tau)= (1-p) S_{ul}+\pc \Bef,
\nonumber \\
&&p=\delta+(1-\delta)\left[\frac{1}{2}K_2(\beef,\tau)+
\frac{1}{2}K_2(\beef,2\tau_0-\tau)\right],
 \\
&&
\pc=\delta\left[1-\frac{1}{2}L_2(\beef,\tau)-\frac{1}{2}L_2(\beef,2\tau_0-\tau)
\right], \nonumber \eeq where $2\tau_0=\int \alw {\rm d} z$ is the
total slab optical depth, $S_{ul}= (\frac{n_l}{n_u}
\frac{g_u}{g_l}-1)^{-1}$ is the line source function (all
intensities are measured in units $2h\nu_{ul}^3/c^2$), $\nuij$ is
the transition frequency, $n_u$ and $n_l$ are the fractional level
populations, $g_u$ and $g_l$ are the statistical weights for the
upper and lower levels, respectively, $\Bef$ is the dust source
function (Planck function), $\beta=\ald/\alw$ is the ratio of the
absorption coefficients due to the dust and the line,
$\delta(\beta)$ is the probability per single interaction that the
line photon will be absorbed by dust, $K_2$ is the probability
that the line photon will escape from the slab and $L_2$ is the
probability that the photon emitted by the dust will escape from
the slab. We use approximations of $\delta$, $K_2$, and $\pc$ from
BP04. In most of our calculations we take $\tau=\tau_0$ (i.e. slab
center). The line absorption coefficient is \be
 \alw=\lambdaij^3 \gi \Aij \Delta \nij \Nw \frac{\nuij}{8 \pi c \Delnud}.
\ee Here,  $\Delta \nij=\nj/\gj-\ni/\gi$,
$\Delnud/\nuij=(2kT/18\mpr c^2)^{1/2}$ is the Doppler width,
$\Aij$ is the Einstein coefficient,  $T$ is the gas kinetic
temperature, $\mpr$ is the proton mass and $\lambda_{ul}=c/\nuij$.

\subsection{Masing transitions}
\label{sec:masing}

Transitions with inverted level populations (masing lines) need a
different treatment. We neglect the dust influence on masing
transitions, i.e. assume $\beta=\delta=p^c=0$ (e.g. $|\beta| \sim
10^{-4}$ for the $6_{16} \rightarrow 5_{23}$ transition).  The
mean intensity of the radiation in the $ul$-line in the slab
center is
%\citep[e.g.][]{AH65, dJ73}
\begin{eqnarray}
J(\tau_x)\simeq  S_{ul} \int_{0}^{1} \, [1-\exp(-\tau_x  / \mu)]
\; {\rm d} \mu, \label{eq:integral}
\end{eqnarray}
where $\mu$ is the cosine of the angle between the direction of
propagation and the outward normal, $\tau_x=\tau_{\rm m}\phi_{x}$,
$\tau_{\rm m}=\sqrt{\pi}\tau_c = \alw^{\rm m} H$ is  the optical
depth in the maser line, $\tau_c$ is the optical depth in the line
center, $\alw^{\rm m}$ is the  maser absorption coefficient,
 $x=(\nu-\nuij)/\Delnud$ is the frequency within the spectral line
in thermal Doppler units, $\phi_{x}=\pi^{-1/2}\exp(-x^2)$ is  the
Doppler profile. In the case of inversion, the optical depth is
negative and the mean intensity grows to infinity. This happens
because the path length parallel to the boundary of the slab
($\mu=0$) becomes infinite. In a real astrophysical system such a
singularity  does not appear because the coherent amplification
length is limited by the size of the maser source (or velocity
gradients). If we approximate the geometry by a cylinder with
half-height $H$ and radius $L$, Eq.~(\ref{eq:integral}) can be
rewritten in the form
\begin{eqnarray}
J(\tau_x)&\simeq&  S_{ul} \int_{0}^{\mu_{\min}} \, [1-\exp(-\tau_x
/ \mu_{\min})] \; {\rm d} \mu
\nonumber \\
&+& S_{ul} \int_{\mu_{\min}}^{1} \, [1-\exp(-\tau_x  / \mu) ] \;
{\rm d} \mu, \label{eq:integral1}
\end{eqnarray}
where $\mu_{\min}=H/L$. The first integral on the right hand side
is $\mu_{\min}[1-\exp(-\tau_x  / \mu_{\min})]$, while the second
one is approximately $[ 1- \mu_{\min} -\mu_{\min}^2/
|\tau_c|\exp(-\tau_x  / \mu_{\min})]$, when $|\tau_x  /
\mu_{\min}| \gg 1$. Therefore, for large $|\tau_x  / \mu_{\min}|$
the mean intensity in the maser line is
\begin{eqnarray} \label{eq:JStau}
J(\tau_x)\simeq S_{ul} [1-\mu_{\min}\exp(-\tau_x  / \mu_{\min}) ].
\end{eqnarray}
At large $|\tau_x  / \mu_{\min}|$,  $J_x$ grows and the masing
transition starts affecting level populations (the maser starts to
saturate). The critical value of $\tau_{\rm m}$ at which the maser
becomes saturated depends on the geometry of the maser source, and
thus on the value of $\mu_{\min}$. The case of $\mu_{\min}=1$
corresponds to an almost spherical maser, while $\mu_{\min} \ll 1$
corresponds to a flat configuration, e.g. a shock wave or an
accretion disk.

%One should mention, that small inaccuracies in the description of
%the  mean intensity in the masing lines $J(\tau_x)$ and the choice
%of $\mu_{\min}$ are not important for calculation  of unsaturated
%as well as fully saturated masers. Accurate description of the
%maser intensity is important only for the transition from one
%regime to another, which  goes very fast (with  changing of
%$\tau_{\rm m}$), because  $J(\tau_x)$ is an exponential function
%of $\tau_{\rm m}$  (see Eq.~\ref{eq:JStau}).

The line-averaged intensity $\overline{J}=\int \phi_{x} J(\tau_x)
\rmd x$ in maser lines is given by Eq.~(\ref{eq:J_new}), where $K_2$
should be replaced by
\be
K_2^{\rm m}=\int_{-\infty}^{\infty} \phi_{x} \mu_{\min}
\exp(-\tau_x  / \mu_{\min}) \rmd x.
\ee
One can show that
\be
K_2^{\rm m} \simeq \frac{ \mu_{\min} \exp(|\tauc|/\mu_{\min}) } {
|\tauc| \sqrt{\pi {\rm ln} |\tauc|}}, \;\;\; {\rm if}\;\;
 |\tau_{\rm c}| \rightarrow \infty ,
\ee while for  $|\tau_{\rm c}| \rightarrow 0$,  $K_2^{\rm m}$ must
smoothly approach $K_2$ for non-masing transitions
\citep{HM79,BP04}. We propose here a formula that connects these
two asymptotes (it generalizes Eq.~[18] in BP04 to
$\mu_{\min}<1$): \be \label{eq:K2m} K_2^{\rm m}= \frac{\mu_{\min}
\exp(|\tauc|/\mu_{\min})+ 1 - \mu_{\min} +1.18|\tauc| }{1+|\tauc|
[\pi\ln(1+|\tauc|)]^{1/2}} . \ee The smoothness of $K_2$, when
$\tau$ passes through zero, allows us to account for the effect of
maser saturation.

To characterize the maser strength we use the optical depth
$\tau_{\rm m} \equiv \alw^{\rm m} H$ determined along the  axis of
the cylinder. This quantity characterizes only local conditions in
the slab center. With increasing $H$, the maser may stop operating
in the slab center, reducing such computed $\tau_{\rm m}$. This,
however, just means that maser may operate more efficiently away
from the central plane closer to the slab boundary.

For a cylindrical maser, the actual observed intensity is larger
along the radius of cylinder than along its axis. To estimate the
maximum power of unsaturated cylindrical masers, one should use
the corresponding optical depth: \be \label{eq:taumax}
\tau_{\max}= (2L/H) \tau_{\rm m}/\sqrt{\pi} =  2L\ \alw^{\rm
m}/\sqrt{\pi} \approx \tau_{\rm m}/\mu_{\min}. \ee
  If the maser is unsaturated (i.e. the maser radiation
does not affect the level populations), the absorption coefficient
in a masing line does not depend on $\mu_{\min}$. However, the
optical thickness along the line of sight and, therefore, the
maser luminosity do depend on $\mu_{\min}$, which is determined by
the geometry of the source. For simplicity of calculations, we
assume $\mu_{\min}=1$ everywhere except an application in
Sec.~\ref{sec:v778cyg}.

\subsection{Population balance}
\label{sec:pop_bal}

Using the solution of the RTE in the escape probability
approximation (\ref{eq:J_new}), the statistical balance equations
take the form (see e.g. BP04)
\beq \label{eq:balafinal} &
&\sum_{\low > \up} \Aji \left( \pji \nj +\pdji \Bef_{lu}
\gj \Delta \nij \right)  \nonumber \\
& - & \sum_{\low<\up} \Aij \left(  \pij \ni + \pdij \Bef_{ul}
\gi \Delta \nji  \right)        \\
& = & \sum_{\low \ne \up} \left(  \Cij  \ni -  \Cji  \nj \right) ,
\quad u=1,2,\dots,M-1, \nonumber \eeq with the normalizing
condition $\sum_{\up} \ni=1$. The Einstein $A$-coefficients
corresponding to the rotational transitions of the ground state
are taken from \citet{CV84b}. Collisional de-excitation rates with
hydrogen, $\Cij$, corresponding to the rotational transitions of
the ground state are from \citet{GM93}, while the excitation rates
are computed using the detailed balance condition. We take into
account the first 45 rotational levels of the ground vibrational
state and 45 rotational levels of the first-excited vibrational
state (010) of ortho- \water\ molecules \citep{To91} and all
possible radiative and collisional transitions between them. We
recalculate the Einstein coefficients of the rotational
transitions in the first-excited  vibration state $A_{ul}^{\rm E}$
and of the vibration-rotational transitions
 $A_{ul}^{\rm EG}$ from the Einstein coefficients of the rotational transitions in the
ground vibrational state $A_{ul}^{\rm G}$  \citep{De77}
\begin{eqnarray}
&&A_{ul}^{\rm E}= A_{ul}^{\rm G} \left( \frac{\nu_{ul}^{\rm
E}}{\nu_{ul}^{\rm G}} \right)^3,
\\
&&A_{ul}^{\rm EG}= A_{ul}^{\rm G} \left(
\frac{\mu'}{\mu_0}\right)^2
 \left( \frac{\nu_{ul}^{\rm EG}}{\nu_{ul}^{\rm G}} \right)^3,
\end{eqnarray}
where
 $\mu'$ is the transition dipole moment between the
vibrational states and $\mu_0$ is the intrinsic dipole moment
($[\mu'/\mu_0]^2=0.005$), $\nu_{ul}^{\rm G}$ is the frequency of
a  rotational transition in the ground vibrational state,
$\nu_{ul}^{\rm EG}$ is the frequency of a vibration-rotational
transition, $\nu_{ul}^{\rm E}$ is the frequency of a  rotational
transition in the first-excited vibrational state. Since the
probabilities of the collisional transitions between vibrational
states $C^{\rm EG}_{ul}$  and of the collisional rotational
transitions in the first-excited vibrational state $C^{\rm
E}_{ul}$ are not known,
 we assume, following  \citet{De77}, that
\be
C^{\rm EG}_{ul}= \chi C_{ul}^{\rm G},
\quad C^{\rm E}_{ul}=0,
\ee
where $C_{ul}^{\rm G}$ is the probability of the rotational
 collisional transitions in the ground vibrational state,
$\chi = 10^{-3}$ is a fixed parameter.

%One should note that the final system of equations mathematically
%does not differ from  that obtained by BP04 for the case of the
%mixture of the  gas with the grains  of one type and size.
%Therefore,  the calculation method used in that paper can be
%directly applied to the multi-dust problems.

The statistical balance equations (\ref{eq:balafinal}) together
with Eqs.~(\ref{eq:J_new}) for escape probabilities are solved by
standard Newton-Raphson method. Our escape probability method is
rather stable and allows us to account for the effect of
saturation in any masing line, which is very difficult (or even
impossible) to implement   in an accelerated lambda iteration
method \citep{YF97}.

\subsection{Dust properties}
 \label{sec:dust}

We consider the following types of dust
\begin{enumerate}
\item amorphous water  ice \citep{HS93},
\item crystalline water ice \citep{Be69},
\item astronomical silicate \citep{LD93},
\item circumstellar silicate \citep{DP95},
\item graphite \citep{LD93},
\item amorphous carbon \citep{RM91}.
\end{enumerate}
Water ice is called  `ice' hereafter.

\begin{figure*}
\centerline{\epsfig{file=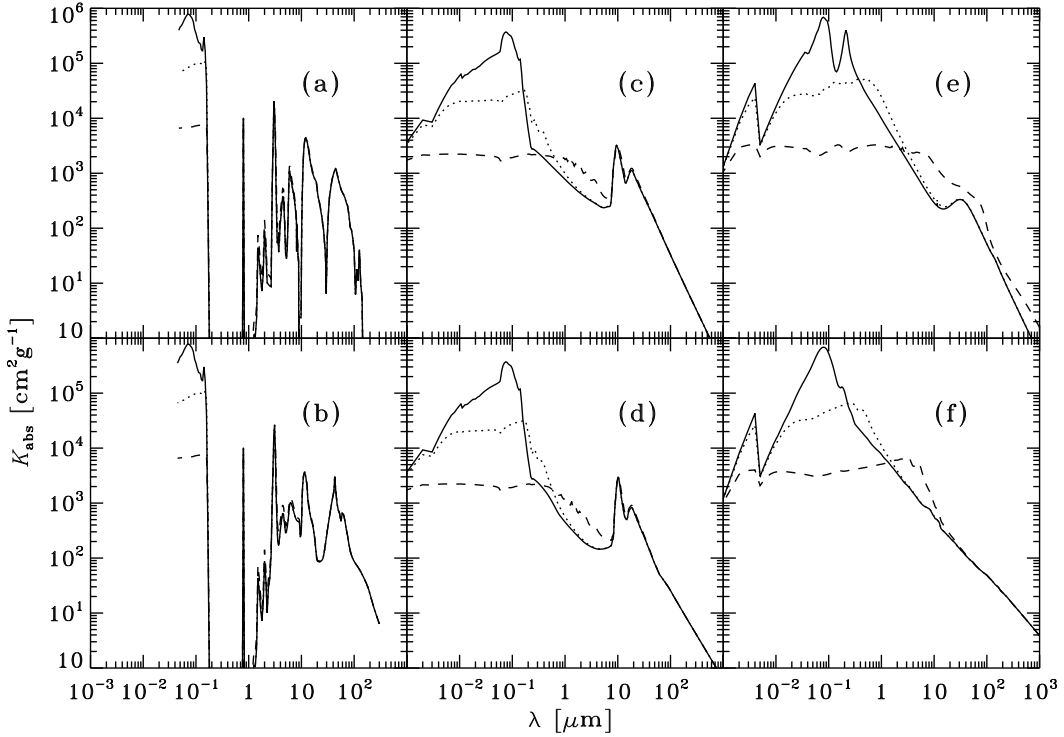,width=14.cm}} \caption{ Dust
absorption coefficient per unit mass, as a function of the
wavelength in the case of spherical dust grains of radius $a=0.01$
$\mum$ (solid curves), $a=0.1$ $\mum$ (dotted curves) and $a=1$
$\mum$ (dashed curves) for  (a) amorphous water ice \citep{HS93},
(b) crystalline water ice \citep{Be69},  (c) astronomical silicate
\citep{LD93}, (d) circumstellar silicate \citep{DP95},  (e)
graphite \citep{LD93}, (f) amorphous carbon \citep{RM91}. }
\label{fig:Kabs}
\end{figure*}

The dust optical properties can be characterized by the absorption
coefficient per unit mass
$\Kabsnu=3 Q_{\lambda}/ 4 \overline{\rho} a  $, where $Q_{\lambda}$  is
the absorption efficiency, $\overline{\rho}$ is the density of the
dust material: 1  $\rm g\,cm^{-3}$  (water ice), 2.26  $\rm
g\,cm^{-3}$ (graphite), 1.85  $\rm g\,cm^{-3}$ (amorphous carbon),
3.3  $\rm g\,cm^{-3}$  (silicates).
 We present $\Kabsnu$  for different dust types and grain sizes
 as  a function of the wavelength in Fig.~\ref{fig:Kabs}.
 Because of the absence of the
experimental data on the optical constants for the amorphous ice
at wavelength $\lambda < 1$ $\mum$, we use data for the
crystalline ice in this wavelength range instead.

We consider either the fixed size of the dust grains or a
power-law distribution of the form \be \label{eq:distr} {\rm
d}n(a) \propto a^{\gamma} {\rm d}a, \quad a_{\rm min}\leq a \leq
a_{\rm max}, \ee with $a_{\rm min}=0.01$ $\mum$, $a_{\rm max}=1$
$\mum$, and  $\gamma=-3.5$ \citep{MR77,Ju96}. In our calculations
we assume the standard dust mass-fraction for interstellar medium
$\fd=10^{-2}$. Since ${\rm d}\fd=\frac{4 \pi}{3}  a^{3}
\overline{\rho} {\rm d}n(a)/(2m_{\rm p} \Nh)$, where
$\overline{\rho}$ is the density of the dust material, one can
write the distribution function for the dust mass-fraction \be
\label{eq:fd_dist} {\rm d}\fd(a) = \fd \frac{  (\gamma+4)}{a_{\rm
max}^{\gamma+4} - a_{\rm min}^{\gamma+4}} a^{\gamma+3}\ {\rm d}a .
%\quad \fd= \int_{a_{\rm min}}^{a_{\rm max}}{\rm d}\fd.
\ee

The dust absorption coefficient is given by $\ald=\Kabs\rhod$,
where $\rhod=\fd 2\mpr \Nh$ is the dust density. If dust of two
different types is present, the dust absorption coefficient is
given by the sum \be \label{eq:be_ef} \ald= \aldc+\aldh =  2\mpr
\Nh
  \left[ \Kabsc \fdc + \Kabsh \fdh \right], \ee
where  $\fdh$ and $\fdc$ are the mass fractions of different dust
types (the total dust  mass fraction is $\fd=\fdh+\fdc$). The dust
source function is then \be \label{eq:B_ef} \Bef= \left[ \aldc
B(\Tdc)+ \aldh B(\Tdh)\right]/\ald , \ee where $\Tdh$ and $\Tdc$
are the corresponding dust temperatures. For the grain size
distribution (\ref{eq:fd_dist}), the  dust absorption coefficient
and the source function are transformed to \beq \ald&=&  2\mpr \Nh
\int  K_{\rm abs} (a) \, {\rm d} \fd , \\
\Bef &=& \frac{ \int  K_{\rm abs} (a)
B[\Td(a)]\,  {\rm d}\fd} {\int  K_{\rm
abs} (a)\, {\rm d} \fd}.
\eeq

We assume that the medium is optically thin to the absorption by
dust,
 \be \label{eq:tau_d} \tau_{\rm d}= 2 \mpr  \fd \Nh
\Kabsave(\Tst) H
%\\
%&=& 3 \left(\frac{\fd}{10^{-2}} \right)
%\left(\frac{\Kabsave(\Tst)}{10^{3} \rm cm^2 g^{-1}} \right)
%\left(\frac{\Nh}{10^9\rm cm^{-3}} \right)\left( \frac{H}{10^{14}
%\rm cm} \right)
\lesssim 1,
\ee
%\nonumber
%\end{eqnarray}
where $\tau_{\rm d}$ is the dust optical depth,
$\Kabsave$ is the Planck-averaged absorption coefficient.

\subsection{Dust cooling and heating}
\label{sec:dust_heat_cool}

\begin{figure}
\centerline{\epsfig{file=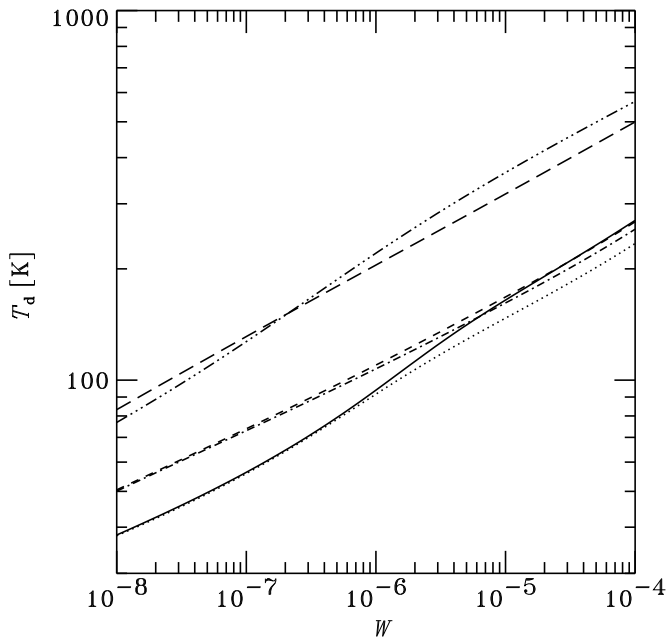,width=8.5cm}}
\caption
{ Dust temperature  for  dust grains of radius $a=0.01$ $\mum$ for
amorphous ice (dotted curve), crystalline ice (solid curve),
astronomical silicate (dashed curve), circumstellar silicate
(dot-dashed curve), graphite (triple-dot-dashed curve), amorphous
carbon (long-dashed curve), as a function of the dilution factor
for $\Tst=3000$ K. } \label{fig:dust_temp}
\end{figure}

The main mechanism for dust heating is provided by the stellar
radiation \be \label{eq:Q+star} q^+_{\rm *} =
%4 \pi W \int{ \alpha_{{\rm d,}\nu} B_{\nu} (\Tst)  d \nu}=
%4 \pi W \Nd A_{\rm d} \overline{Q}(\Tst) \sigma \Tst^4,
W 8 \mpr \Nh \fd   \Kabsave(\Tst)  \sigma \Tst^4 , \ee where
$W=\frac{1}{4}  \left(\frac{R}{D}\right)^2$ is the dilution
factor, $R$ is the stellar radius, and $D$ is the distance to the
star. The  dust cools by its own radiation with the corresponding
rate \be \label{eq:Q-rad} q^-_{\rm rad} =
%4\pi \int{ \alpha_{{\rm d,}\nu} B_{\nu} (\Td)  d \nu}=
%4  \pi \Nd A_{\rm d} \overline{Q}(\Td)  \sigma \Td^4 .
 8 \mpr \Nh \fd  \Kabsave(\Td)  \sigma \Td^4 .
\ee
In addition, there is energy exchange  between gas and dust
\citep{GK74}
\be \label{eq:Q-gas} \label{eq:Q+gas}
q_{\rm gas} = \Nd A_{\rm d} \Nh V_{\rm t} k (T-\Td) =
 \frac{3 \mpr }{2\overline{\rho} } \frac{\fd}{a} \Nh^2 V_{\rm t} k (T-\Td)
, \ee where $\Nd$ is the concentration of the dust grains,
$\Ad=\pi a^2$ is their geometrical cross sections, and $V_{\rm t}$
is the hydrogen thermal velocity. The energy exchange  is small
compared to radiative cooling and heating if $\Nh \lesssim 10^{9}$
cm$^{-3}$ and can be neglected. Absorption of the near-infrared
photons from $\rm H_2 O$ and $\rm CO$ present in the dusty medium
gives an even smaller contribution to the energy exchange. The
thermal balance for the dust is then transformed to a standard
form \be \label{eq:th_bal_dust} q^+_{\rm *} \simeq q^-_{\rm rad} .
\ee The  dust temperatures  depend only on the stellar
temperature, the dilution factor (approximately as $W^{1/5}$) and
the dust optical properties (see Fig.~\ref{fig:dust_temp}).

\subsection{Gas heating and cooling}
\label{sec:gas_heat_cool}

The main heating and cooling mechanisms determining the gas
kinetic temperature are the dust-gas collisions, photoelectric
effect on grains \citep{dJ77,TH85,Gr94}, cosmic rays \citep{GL78},
viscous stresses \citep{SS73} and  radiation of $\rm H_2O$ and
$\rm H_2$ molecules \citep[e.g.][]{GS76}. The heating (cooling)
rate of the gas per unit volume $(\rm erg \, cm^{-3} s^{-1})$ due
to collisions  with hotter (cooler) dust  is given by
Eq.~(\ref{eq:Q-gas}): \be \label{eq:gasdust} q_{\rm dust} =-q_{\rm
gas}. \ee The rate of gas heating due to cosmic rays and viscous
stresses is several orders of magnitude smaller than that due to
the dust-gas collisions, while heating of the gas due to the
photo-effect is important when the stellar radiation is strong in
the UV range, and can be neglected for late-type stars. Moreover,
in spite of the fact that hydrogen is the most abundant molecule
in the oxygen-rich material, the rate of gas cooling due to $\rm
H_2$  \citep{HD80} is much smaller than that due to the dust-gas
collisions.

\begin{figure}
\centerline{\epsfig{file=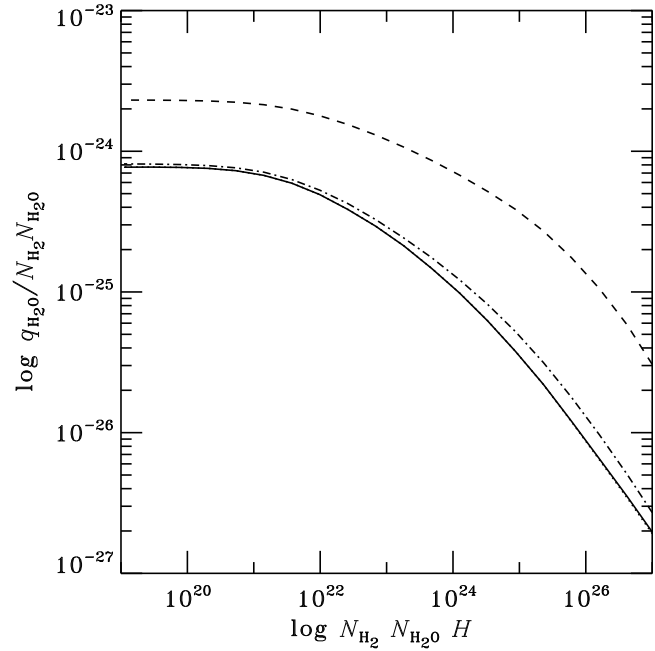,width=8.5 cm}}
\caption {
Water cooling efficiency (erg cm$^3$ s$^{-1}$) as a function of
$\Nh \Nw H$ in the case of the water-hydrogen mixture, including
the first vibrational level in water molecule: with silicate dust
(solid curve) and  without dust (dashed curve); taking  into
account only ground vibrational state: with silicate dust (dotted
curve) and  without dust (dot-dashed curve). The dust and gas
temperatures are 200 K, the grain size is 0.01 $\mum$, dust-to-gas
mass ratio is $\fd=10^{-2}$, the slab half-thickness is
$H=10^{14}$ cm. } \label{fig:cool_rate}
\end{figure}

Next to $\rm H_2$ the most abundant molecule in oxygen-rich matter
is probably the $\rm H_2O$ molecule \citep{GS76,Gr94}.
\citet{NL95} obtain cooling functions for five molecules and two
atomic species that are potentially important coolants ($\rm H_2$,
$\rm H_2O$, CO, $\rm O_2$, HCl, C, and O), and show that, for a
temperature of about 200 K and hydrogen concentration $
10^{6}\lesssim \Nh \lesssim 10^{9}$ cm$^{-3}$, the gas  cools
mostly by $\rm H_2O$ molecules. The heat loss rate of the gas due
to radiative cooling by $\rm H_2O$ is \be \label{eq:H20_cool}
q_{\,\rm H_2O}=\sum{\Nw h \nu_{ul} ( n_l C_{lu} - n_u C_{ul})},
\ee where the sum is over all  collisional transitions in the
water molecule, $C_{lu}$ and $C_{ul}$ are the rates of the
collisional excitation and de-excitation. In principle, the value
of  $q_{\rm H_2O}$ can be negative, which means that the radiative
cooling in water lines is less than the heating due to absorption
of dust radiation.

In order to test our code, we compute the cooling rate due to
water lines.  Fig.~\ref{fig:cool_rate} shows the dependence of
$q_{\rm H_2O}/(\Nh \Nw)$   on $\Nh \Nw H$. First, we  take into
account only the ground vibrational state in the water molecule
and then include the first-excited vibrational level. Our results
are similar to those obtained  by  \citet{NM87}, who accounted for
the first 179 rotational levels of the ground vibrational state.
The small difference can be explained by the different number of
rotational levels included. In the presence of dust with $\Td=T$
the cooling is reduced because dust grains trap  part of the
radiation from water molecules  in the medium. If one includes the
first-exited vibrational state in the water molecule, the cooling
rate becomes larger because of additional vibration-rotational
collisional transitions and rotational transitions in the
first-exited vibrational state, but with dust there is no
difference.

\begin{figure}
\centerline{\epsfig{file=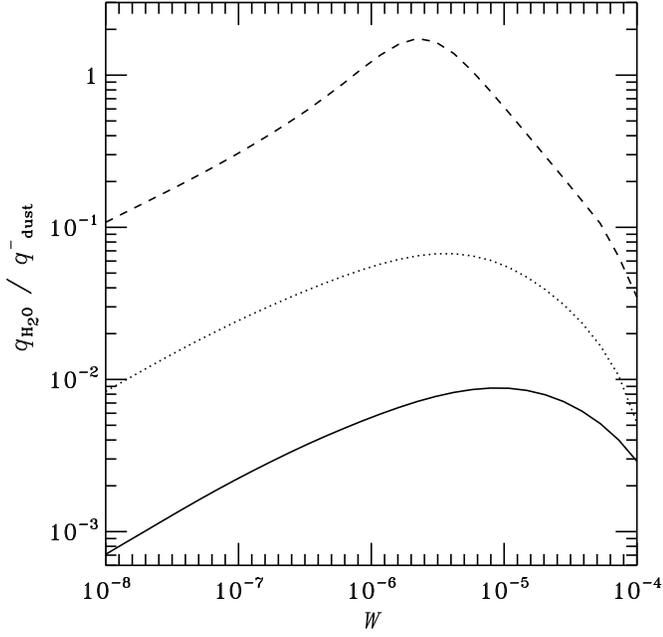,width=8.5 cm}}
\caption { Dependence of the ratio  $q_{\rm H_2O}/q^-_{\rm dust}$
 on dilution factor $W$ for a circumstellar silicate
and amorphous carbon dust mixture at the fixed values of slab
half-thickness $H=10^{14}$ cm, the gas concentration $\Nh=10^{8}$
cm$^{-3}$, water mass fraction $\fw=10^{-4}$, dust mass fraction
$\fd=10^{-2}$, temperature of the star $\Tst=3000$ K and different
grain size $a=0.01$ $\mum$ (solid curve), 0.1 $\mum$ (dotted
curve), 1 $\mum$ (dashed curve), assuming the same mass fraction
of the two dust types. } \label{fig:cooling}
\end{figure}

A rather general form of the thermal balance equation for the gas
is \be \label{eq:th_gas_bal}
 q _{\rm dust}=   q_{\rm H_2O}.
\ee The thermal balance for the mixture of the gas and  different
dust types (noted by index $ i$) is \be \label{eq:th_gas_bal_new}
\sum{ N_{{\rm d},i} A_{{\rm d},i} \Nh V_t k ( T_{{\rm d},i} - T)}=
q_{\rm H_2O}. \ee The gas temperature can be obtained by
simultaneously solving this equation with the  population balance
equation for a water molecule (see section \ref{sec:pop_bal}).
Such a system of equations is non-linear and is solved by
iterations. If the radiative cooling by $\rm H_2O$  is small, the
gas temperature is expressed by a simple relation \be
\label{eq:T_gas_new} T =\frac{ \sum_{i} N_{{\rm d},i} A_{{\rm
d},i} T_{{\rm d},i}}{ \sum_{i} N_{{\rm d},i} A_{{\rm d},i}} =
\frac{ \sum_{i} f_{{\rm d},i} T_{{\rm d},i} / \overline{\rho}_{i}
a_i }{ \sum_{i} f_{{\rm d},i}  / \overline{\rho}_{i} a_i  } . \ee
In the case of the grain size distribution, the energy balance for
the gas is \be \label{eq:dist_bal} \frac{3 \mpr}{2
\overline{\rho}} \Nh^2V_t k \int_{a_{\rm min}}^{a_{\rm max}}
[\Td(a)-T] a^{-1} {\rm d}\fd= q_{\rm H_2O}. \ee When various dust
types are present, some can cool the gas while others can heat it.

We illustrate the effect of water cooling with a simple example of
two dust types mixed with hydrogen and water. We consider equal
mass fractions of circumstellar silicate and amorphous carbon with
total $\fd=10^{-2}$. Silicate is cooler, while carbon is hotter.
It is clear from Eq.~(\ref{eq:dist_bal}) that small grains are
more effective in heating (and cooling) the gas.
Fig.~\ref{fig:cooling} compares the water cooling rate to that due
to collisions with silicates for different grain sizes $a=0.01$,
0.1, and 1~$ \mum$ as a function of the dilution factor. We see
that the radiative cooling rate due to $\rm H_2O$ is about two
orders of magnitude smaller than the cooling due to collisions
with 0.01 $ \mum$ grains, while the efficiency of these two
mechanisms is comparable for 1~$\mum$ grains.

\begin{figure}
\centerline{\epsfig{file=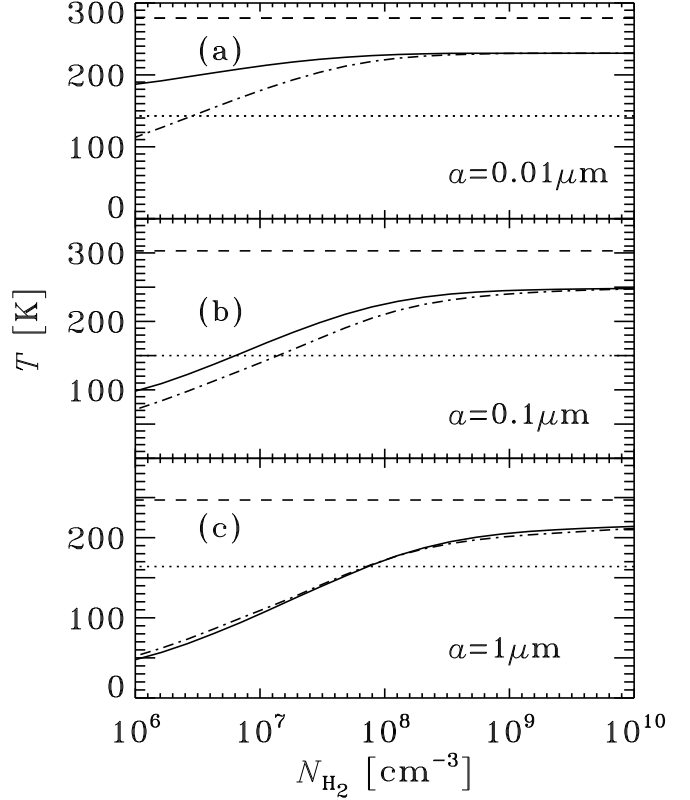,width=8.5 cm}}
\caption {Gas temperature as a function of hydrogen concentration
in a slab of half-thickness $H=10^{14}$ cm in the
radiation field with $W=5 \times 10^{-6}$ and $\Tst=3000$ K and
 different grain sizes {\bf (a) } $a=0.01$ $\mum$, {\bf  (b) }
0.1 $\mum$, {\bf (c) } 1 $\mum$.
Water fraction of $\fw=10^{-4}$ (solid curves) and
$\fw=10^{-3}$ (dot-dashed curves) is assumed.
Dust of two types, silicate and amorphous carbon,
is present with equal mass fraction and total $\fd=10^{-2}$.
Dust temperatures of the silicate  and amorphous carbon are shown by
dotted and dashed lines, respectively.
  } \label{fig:cool_Nh}
\end{figure}

The effect of radiative  cooling by $\rm H_2O$ on the gas
temperature does not depend only on the size of the dust grains,
but also on the hydrogen concentration. This is because the
heating and cooling rates have different dependencies  on $\Nh$:
$q_{\rm dust}(\Nh) \propto \Nh^2 \fd$ (see Eqs.~[\ref{eq:Q-gas},
\ref{eq:dist_bal}]) and $q_{\rm H_2O}(\Nh) \propto (\Nh \Nw)^{1/2}
\propto\Nh \fw^{1/2}$ (see Fig.~\ref{fig:cool_rate}). In
Fig.~\ref{fig:cool_Nh} we illustrate the dependence of $T$ on
$\Nh$ for the dilution factor $W=5 \times 10^{-6}$ (where water
cooling is most effective, see Fig.~\ref{fig:cooling}). At high
$\Nh$  the gas temperature takes an intermediate value given by
Eq.~(\ref{eq:T_gas_new}). In the case of 0.01 $\mum$ grains,  $\rm
H_2O$ radiative cooling is negligible for $\Nh \gtrsim 10^{8}$
cm$^{-3}$, while in the case of 0.1 and 1 $\mum$ grains, water
cooling is important even for $\Nh \lesssim 10^{9}$ cm$^{-3}$.
Moreover, for large grains and $\Nh \lesssim 10^{7}$ cm$^{-3}$
water cooling is so effective that the gas becomes cooler than the
cold dust (silicate).

\begin{figure*}
\centerline{\epsfig{file=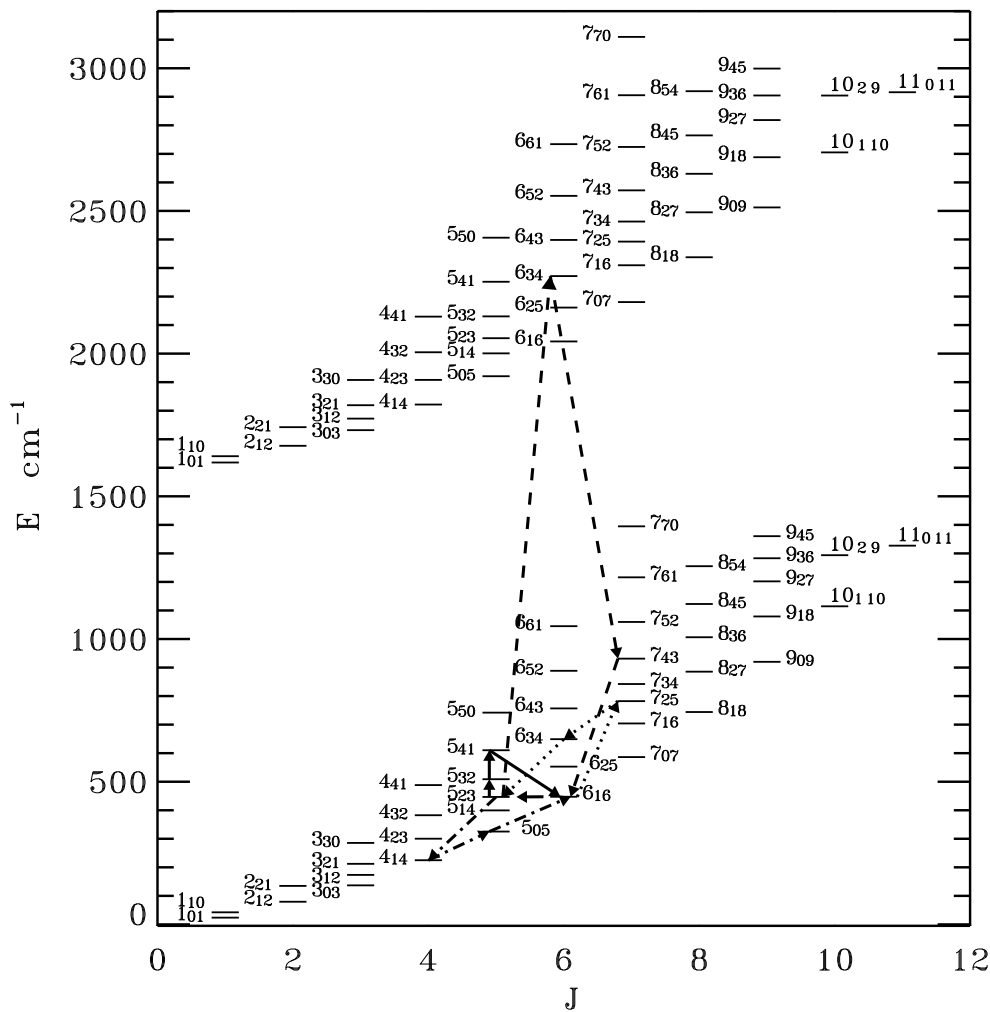,width=13.cm}} \caption {
System of the rotational energy levels of ortho-$\rm H_2O$
molecule in the ground and  first-excited (010) vibrational state.
Arrows show the main cycles of  the maser pumping in the following
cases: Deguchi model (dot-dashed lines), de Jong model (solid
lines), two-dust model (dashed lines). The cycle extinguishing the
maser  effect at large $H$ for the combinations of the ice with
any other dust is shown by dotted lines. } \label{fig:levels}
\end{figure*}

In the winds from late-type stars, the radiation pressure from a star
accelerates the dust which can drift through the gas,
providing an additional source of gas heating \citep[see, for example,][]{GS76}
 \be \label{eq:heat_dyn}
q_{\rm drift} \simeq \frac{1}{2}\Nh 2 \mpr   \sum_{i} N_{{\rm d},i}
A_{{\rm d},i} V_{{\rm d},i}^3 ,
\ee
 where $V_{\rm d}$ is the drift velocity.
This term should be added to the balance equations
(\ref{eq:th_gas_bal}) and (\ref{eq:th_gas_bal_new}).

\section{Results}

\label{sec:results}

\subsection{de Jong model}
\label{sec:dJ}

Let us first discuss how the $6_{16}-5_{23}$ water maser works if
no dust is present in the medium. We will thus have a benchmark to
which we can compare our dusty models. De Jong (1973) considered a
gas cloud consisting of a mixture of water vapor and hydrogen. He
showed that upon approach  to the surface of the cloud the optical
depth in the lines decreases and the thermal equilibrium breaks
down. This occurs at different depths in different lines and over
some range of depths some levels become relatively overpopulated,
leading to population inversion in some transitions. The
collisional rates of the water molecule with hydrogen were not
known at that time, and he used  approximate expressions. We
repeat these calculations using modern collisional rates from
\citet{GM93}. To find the main transitions participating in the
maser pumping  mechanism we compared the population fluxes
$F_{ul}=n_u W_{ul} - n_l W_{lu}$ (where $W_{ul}$ and $W_{lu}$ are
total transition probabilities) for all possible pure rotational
and ro-vibrational transitions of ortho-$\rm H_2 O$ molecule in
the ground and  first-excited vibrational state.

We find that for $\Nh \gtrsim 10^8$ cm$^{-3}$ and $\fw \gtrsim
10^{-5}$ the main cycle of the maser pumping is $5_{23}
\buildrel161\mum\over\longrightarrow 5_{32}
\buildrel99\mum\over\longrightarrow 5_{41}
\buildrel61\mum\over\longrightarrow 6_{16}$ (solid lines in
Fig.~\ref{fig:levels}). The upward transitions $5_{23}\rightarrow
5_{32}$ and $5_{32}\rightarrow 5_{41}$ are dominated by collisions
and  therefore, water molecules are collisionally excited from
$5_{23}$ to $5_{41}$ level through $5_{32}$ level. Radiative
de-excitation from  $5_{41}$ to $6_{16}$ produces the heat sink in
the maser pumping cycle, when radiative rate of this transition is
larger than the collisional one and the optical depth in the
corresponding line is  small. In the remaining range of parameters
$\Nh$ and $\fw$, the main cycle of the maser pumping is $5_{23}
\buildrel45\mum\over\longrightarrow 4_{14}
\buildrel28\mum\over\longrightarrow 7_{07}
\buildrel72\mum\over\longrightarrow 6_{16}$. Water molecules are
radiatively de-excited from the $5_{23}$ to $4_{14}$ level,
collisionally excited to the $7_{07}$ level, and again radiatively
de-excited to $6_{16}$ closing the pumping cycle.
%$4_{14} \rightarrow 7_{07} \rightarrow 6_{16}$  close  the  maser pumping.
The pumping  cycles found by us appear to be different from those
considered by de Jong, probably because of more accurate
collisional rates in our calculations.

\begin{figure}
\centerline{\epsfig{file=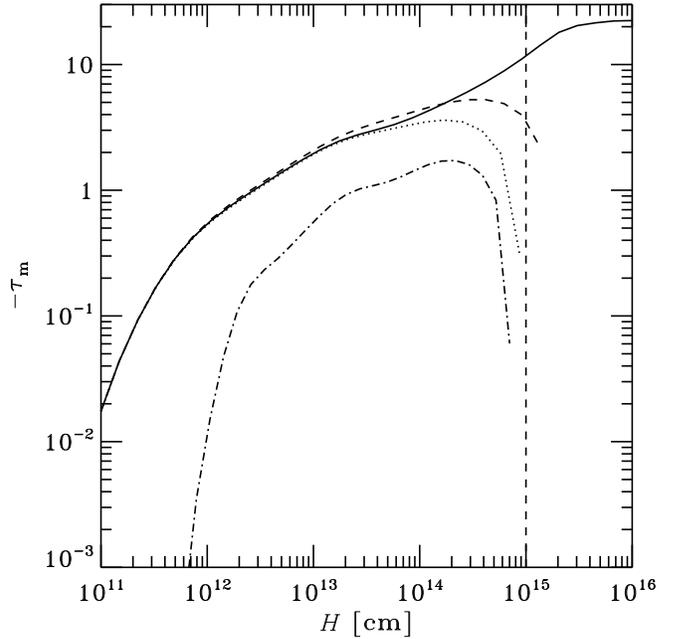,width=8.5 cm}} \caption{ The
optical depth in the  $6_{16}-5_{23}$ maser line as a function of
the half-thickness of the slab for   $\Nh=10^{8}$ cm$^{-3}$,
$\fw=10^{-3}$. The following cases are considered: (1) de Jong
mechanism in the water-hydrogen mixture (without dust) of $T=230$
K (dashed curves); (2) same as previous, but with circumstellar
silicate grains of $a=0.01$ $\mum$ and $\Td=T$   (dotted curves);
(3) gas mixture with amorphous carbon  grains of $a=1$ $\mum$  and
$\Td=279$ K. The gas temperature is calculated self-consistently,
using  thermal balance equation (\ref{eq:th_gas_bal_new})
(dot-dashed curves); (4) circumstellar silicate grains of $a=0.01$
$\mum$ and $\Tdc=143$ K, gas with $T=230$ K, and 0.01$\mum$
amorphous carbon grains with $\Tdh=279$ K (solid curves).
 Dust temperatures  correspond
to the dilution factor $W=5 \times 10^{-6}$ and stellar
temperature $\Tst=3000$ K. We assume  $\fdh=\fdc=\fd/2$,
$\fd=10^{-2}$, and  $\mu_{\min}=1$. On the right hand side of the
vertical dashed lines the dust is optically thick. }
\label{fig:Models}
\end{figure}

In Fig.~\ref{fig:Models} (case 1, dashed curves) we present  the
maser absorption coefficient as a function of the slab
half-thickness. The maser disappears at $H \gtrsim 10^{15}$ cm, if
$\Nh=10^{8}$ cm$^{-3}$, when the probability of photon escape in
$5_{41}\rightarrow 6_{16}$ transition (heat sink) becomes
comparable to the rate of collisions, i.e. $C_{5_{41}\rightarrow
6_{16}} \simeq p A_{5_{41}\rightarrow 6_{16}}$ (see
Eqs.~[\ref{eq:J_new},\ref{eq:balafinal}] and BP04).
%\begin{eqnarray} \label{eq:md}
% 8 \times 10^{-13} \Nh &\simeq& 3 \times 10^{-4} K_2(\tau)
%\nonumber \\
%&=&\frac{3 \times 10^{-4}}{ 1 + \tau [2 \pi \ln (2.13 + \tau^2)]^{1/2}}.
%\end{eqnarray}
For the same water concentration but higher hydrogen density of
$\Nh=10^{9}$ cm$^{-3}$, this condition is satisfied at smaller
line optical depth $\tau$ corresponding to $H \sim 10^{14}$ cm,
where the maser disappears.

\subsection{Influence of dust}

Let us now investigate the influence of the dust on the inversion
of the maser level populations. If dust (we consider 0.01 $\mum$
circumstellar silicate grains) is of the same temperature as the
gas, the maser efficiency drops at large depths (compare case 2,
dotted curves in Fig.~\ref{fig:Models} to case 1, dashed curves).
This happens because dust traps radiation from water molecules in
the medium, while the de Jong mechanism is based on the escape of
photons from the surface.

We have not specified any mechanism of gas heating. In the models
considered above we have assumed certain gas and dust temperatures
without considering energy balance. If there is no relative motion
between the dust and the gas, the only efficient mechanism of gas
heating is the energy exchange with the dust. If only one type of
dust is present, the gas temperature can be at most that of the
dust, and if water cooling is important then it is much smaller.
The hotter dust not only heats the gas, but strongly affects the
maser operation because of the influences of the dust radiation.

\citet{GK74} proposed that the inversion of the maser level
populations $6_{16}$--$5_{23}$ can arise due to the excitation of
the first-excited vibration state (010) of the ortho-$\rm H_2O$
molecule by 6.3 $\mum$ dust radiation, while  the heat sink is
realized through collisions with cooler hydrogen. This model has
been criticized by \citet{De81}, who has pointed out that at high
concentrations needed for the collisional sink, the levels at the
ground vibrational state will be thermalized.

To check this scenario, we considered amorphous carbon grains of
size $a=1$ $\mum$ in the radiation field with stellar temperature
$\Tst=3000$ K and dilution factor  $W=5 \times 10^{-6}$. The dust
temperature $\Td=279$ K  is determined from the thermal balance
equation (\ref{eq:th_bal_dust}). The gas temperature, which is
computed self-consistently from Eq.~(\ref{eq:th_gas_bal_new})
accounting for the water cooling and the heating  by collisions
with the dust, varies from 43 K at $H=10^{11}$ cm to 276 K at
$H=10^{16}$ cm (for $\Nh=10^8$ cm$^{-3}$). The inversion of maser
levels (see case 3, dot-dashed curves in Fig.~\ref{fig:Models}) is
smaller than in previously considered cases and is still produced
by the de Jong mechanism. At small $H$  the radiative cooling by
water is so intense that the gas temperature becomes too low for
maser pumping. In spite of inability of the hot dust to
efficiently pump the maser, the presence of the  hot dust  settles
the problem of the gas heating.

Of course, if there are additional mechanisms of gas heating (e.g.
due to dynamic friction with rapidly moving dust, see
Eq.~[\ref{eq:heat_dyn}]), the gas temperature can exceed the dust
temperature by a large factor. In this situation the maser is very
effective
 (as was discussed for example by \citealt{CK84a,CW95,YF97}; BP04).

\begin{figure}
\centerline{\epsfig{file=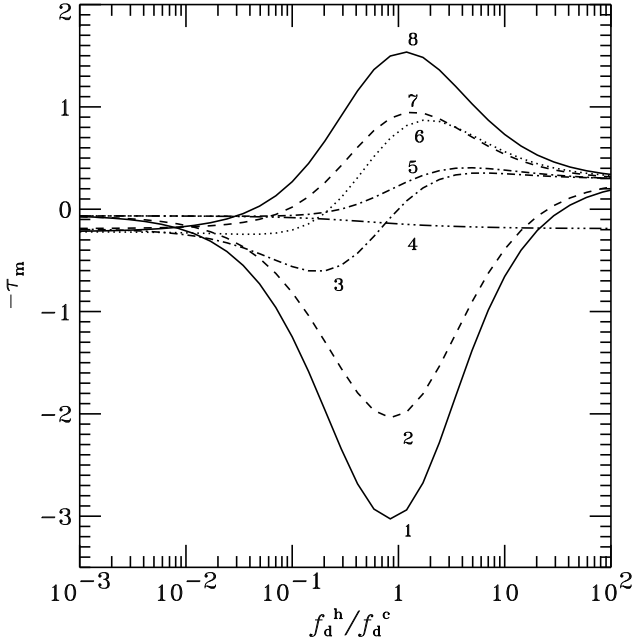,width=8.5 cm}}
\caption { The optical depth $\tau_{\rm m}$ in the $6_{16}
\rightarrow 5_{23}$ transition as a function of the  ratio
$\fdh/\fdc$ for different cold -- hot dust combinations: amorphous
ice and graphite  (1), crystalline ice and graphite  (2),
crystalline ice and amorphous carbon (3), crystalline ice and
circumstellar silicate (4), circumstellar  silicate and graphite
(5), astronomical silicate and graphite  (6), circumstellar
silicate and amorphous  carbon (7), astronomical silicate and
amorphous carbon (8). The parameters are $H=10^{14}$ cm, $W= 5
\times 10^{-6}$, $\Tst=3000$ K and  the grain size  $a=0.01$
$\mum$. } \label{fig:f_hc_maser}
\end{figure}

\subsection{Hot and cold dust pumping mechanism}

The late-type stars often show emission of different types of dust
\citep{DB94}. Thus, it is natural to assume that two dust types
are present simultaneously. Since dust temperatures differ (see
Fig.~\ref{fig:dust_temp}), collisions with one dust can heat the
gas, while with another can cool it. The gas temperature then
takes an intermediate value (see Fig.~\ref{fig:cool_Nh}).

\begin{figure}
\centerline{\epsfig{file=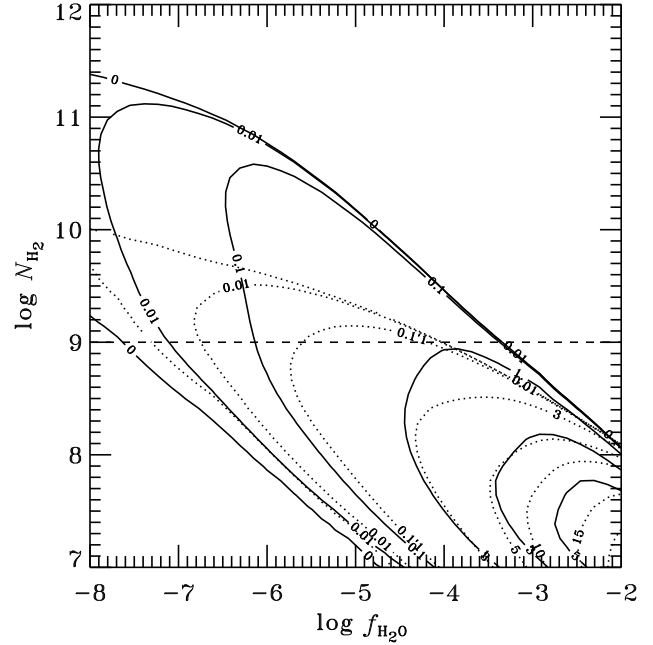,width=8.5 cm}}
\caption { Contour plots of levels of constant the $6_{16}
\rightarrow 5_{23}$ maser optical depth $-\tau_{\rm m}$. Solid
curves correspond to the case of the circumstellar silicate dust
with $\Tdc=143$ K and amorphous carbon with $\Tdh=279$ K; dotted
curves correspond to the case of equal gas and dust temperatures
$T=\Td=230$ K. The size of the dust grains is $a=0.01$  $\mum$ and
the slab half thickness is $H=10^{14}$ cm, the mass fraction of
the silicate and carbon dust is the same
 $\fdc=\fdh=5 \times 10^{-3}$.
On the upper side of the horizontal dashed lines the dust is
optically thick. } \label{fig:Nh_ata_contur}
\end{figure}

What pairs of dust produce the largest inversion? From
Fig.~\ref{fig:dust_temp} we see that a significant temperature
difference can be achieved for dust types where one is graphite or
amorphous carbon and another one is any of the remaining
(silicates or ice). For completeness, we also consider a mixture
of crystalline ice and circumstellar silicate, in spite of the
fact that the temperature difference is not so large.
Fig.~\ref{fig:f_hc_maser} shows the dependence of the optical
depth in the maser line $\tau_{\rm m}$ on the ratio of the
hot-to-cold mass fractions $\fdh/\fdc$ for different dust
combinations. We see  that the strongest maser (i.e. largest
negative optical depth $\tau_{\rm m})$ occurs at a ratio
$\fdh/\fdc \simeq 1$ for  combinations involving graphite (or
amorphous carbon) and silicates. We further assume  the same mass
fraction of the cold and hot dust, $\fdh=\fdc$.

Let us now consider a mixture of 0.01 $\mum$ circumstellar
silicate and amorphous carbon grains. Their temperatures, in the
radiation field with $W=5 \times 10^{-6}$ and $\Tst=3000$ K, are
$\Td^{\rm cs}=143$ K and $\Td^{\rm ac}=279$ K, respectively. The
gas temperature of $T=230$ K is fully determined by the collisions
with dust which are more efficient than water radiative cooling
for small grains (see section \ref{sec:gas_heat_cool}). The
resulting maser strength as a function of the slab thickness is
shown in Fig.~\ref{fig:Models} (case 4, solid curves). We see that
at $H \lesssim 10^{14}$ cm, the inversion is identical to that in
the dust-free case, implying operation of the de Jong mechanism.
At larger $H$, the maser absorption coefficient is almost constant
resulting in a linear increase of the maser optical depth. At even
higher $H \gtrsim 3 \times 10^{15} $ cm the maser saturates (we
used $\mu_{\min}=1$ in these calculations, i.e. almost spherical
maser, see Eq.~\ref{eq:K2m}). Thus, we see that the presence of
two dust types is rather efficient in heating the gas as well as
keeping a large inversion deep in the cloud. For $\Nh H \gtrsim
10^{23}$ cm$^{-2}$ the medium becomes optically thick to
absorption by dust, and a large difference  between  the gas and
dust temperatures is unlikely to be produced in real astrophysical
environments (unless there is additional heating, of course).

\begin{figure}
\centerline{\epsfig{file=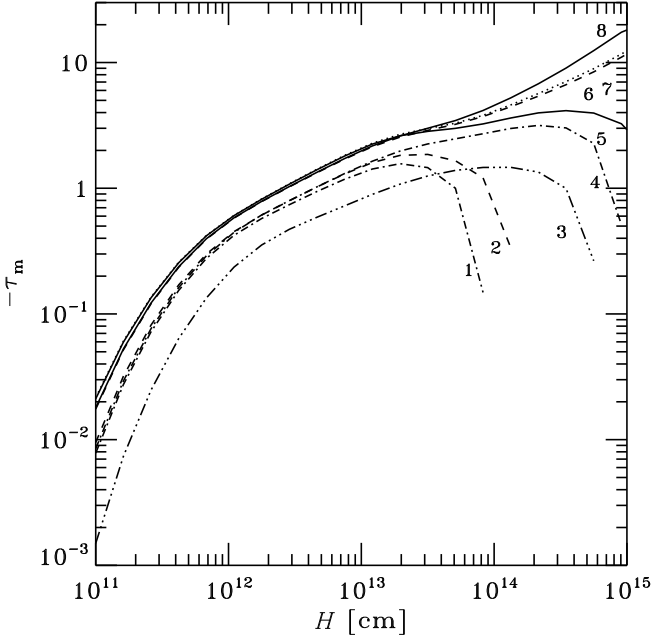,width=8.5 cm}} \caption { The
optical depth in the $6_{16}-5_{23}$ maser line as a function of
the slab half-thickness for the same cold-hot dust combination as
in Fig.~\ref{fig:f_hc_maser} and the following set of parameters:
$a=0.01$ $\mum$, $\Tst=3000$ K, $W=5 \times 10^{-6}$, $\fdh=\fdc=5
\times 10^{-3}$.   } \label{fig:H_maser}
\end{figure}

In Fig.~\ref{fig:Nh_ata_contur} we present contour plots of levels
of the constant $6_{16} \rightarrow 5_{23}$ maser optical depth
$\tau_{\rm m}$ at the plane  hydrogen concentration --
water-to-gas mass ratio. The gas temperature is calculated
self-consistently from the thermal balance equation
(\ref{eq:th_gas_bal_new}). For large $\Nh$ or small water content,
the gas temperature is 230 K, while for smaller $\Nh$ and large
$\fw$, the gas cooling by $\rm H_2O$ is important and $T$
decreases down to 150 K (lower right corner in
Fig.~\ref{fig:Nh_ata_contur}). We assume that the dust
temperatures are constant in the whole range of $\Nh$. However,
for $\Nh \gtrsim 10^9$ $\rm cm^{-3}$ (and our assumed $H=10^{14}$
cm) the medium becomes optically thick to the dust absorption and
therefore the gas and dust temperatures equalize. To investigate
the maser effect in the case of equal gas and dust temperatures,
we also present in Fig.~\ref{fig:Nh_ata_contur} (dotted curves)
the maser optical depth for $T=\Td=230$ K. In that case the
inversion of maser levels exists up to  $\Nh \lesssim 10^{10}$
$\rm cm^{-3}$ and the maser is weak in the range $ 10^{9} \lesssim
\Nh \lesssim 10^{10}$ $\rm cm^{-3}$. Thus, the  $6_{16}
\rightarrow 5_{23}$ maser is mostly effective in the medium with
$\tau_{\rm d } \lesssim 1$ (see Eq.~[\ref{eq:tau_d}]).

\begin{figure}
\centerline{\epsfig{file=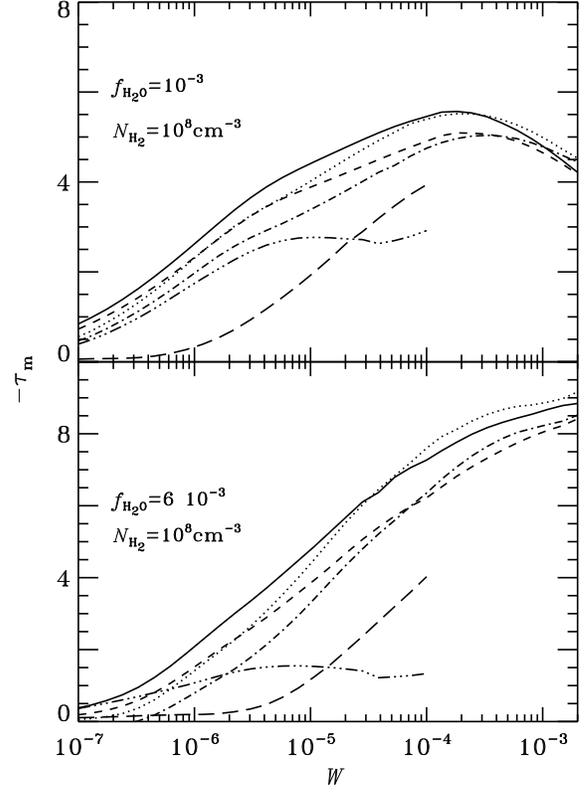,width=8.5cm}}
\caption {
The $6_{16}-5_{23}$ maser optical depth as a function of the
dilution factor  $W$ for different cold--hot dust combinations:
astronomical silicate  and amorphous carbon (solid curves),
astronomical silicate and  graphite (dotted curves), circumstellar
silicate and amorphous carbon  (dashed curves), circumstellar
silicate and   graphite (dot-dashed curves), crystalline ice  and
amorphous carbon (triple-dot-dashed curves), crystalline ice and
circumstellar  silicate (long-dashed curves). The parameters are
$\fdh=\fdc=5 \times 10^{-3}$, $a=0.01$ $\mum$, $H=10^{14}$ cm, and
$\Tst=3000$ K.
Maser efficiency for ice is not computed for $W>10^{-4}$,
because of its evaporation at corresponding temperatures.
} \label{fig:W_maser}
\end{figure}

The $6_{16}-5_{23}$ maser optical depth as a function of $H$ is
shown in Fig.~\ref{fig:H_maser} for the same cold-hot dust
combinations as in Fig.~\ref{fig:f_hc_maser}. We see that  the
maser is less effective for the combinations of ice with other
dust types and  most effective for the combinations of  silicates
with amorphous carbon. The maser works due to the de Jong
mechanism when $H \lesssim 10^{14}$ cm, while for the silicates--
carbon (graphite) mixtures the inversion   exists also at larger
$H$, where  the maser action is controlled by the dust.

\begin{figure*}
\begin{center}
\leavevmode \epsfxsize=8.5cm \epsfbox{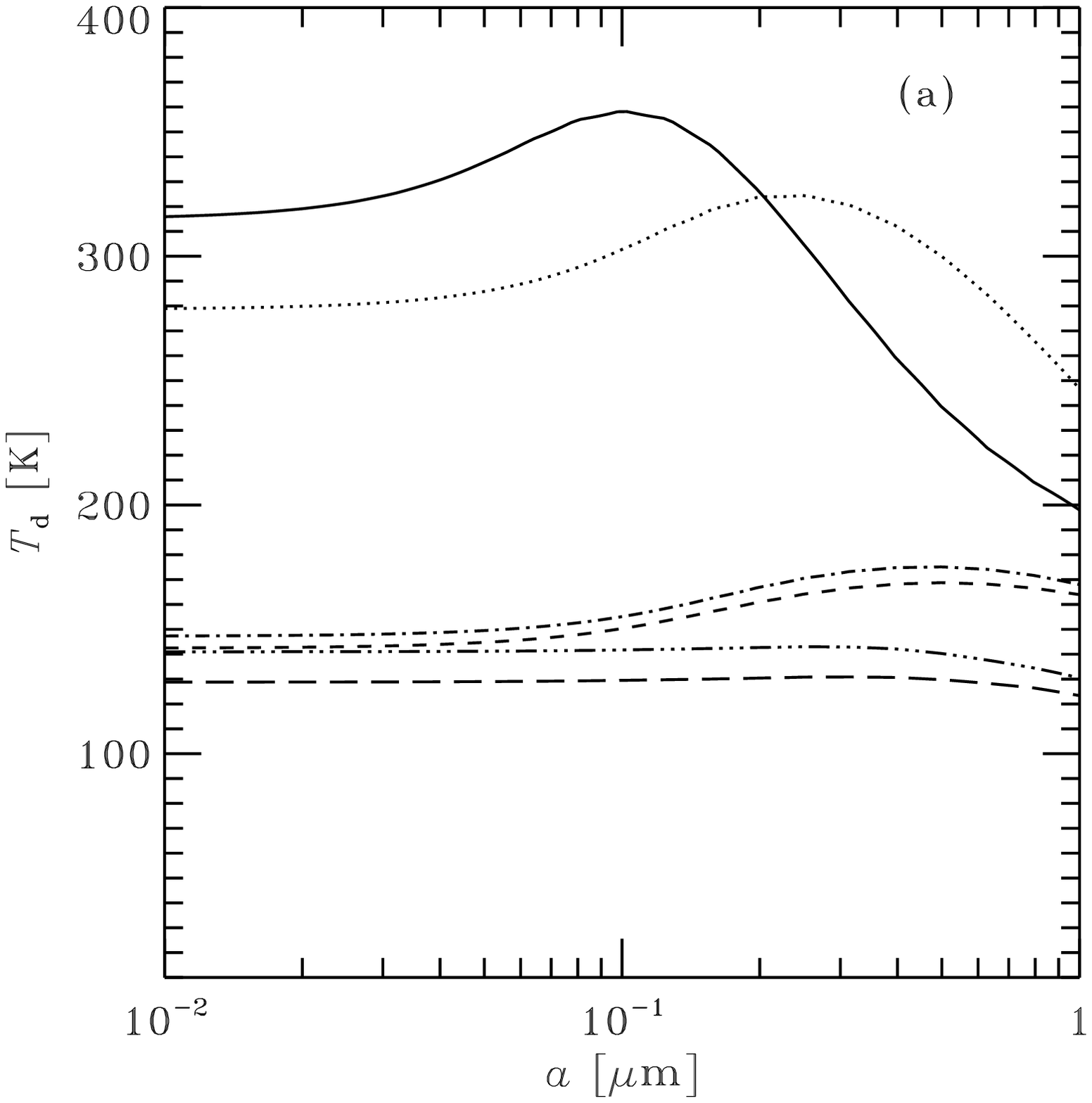}
\epsfxsize=8.5cm \epsfbox{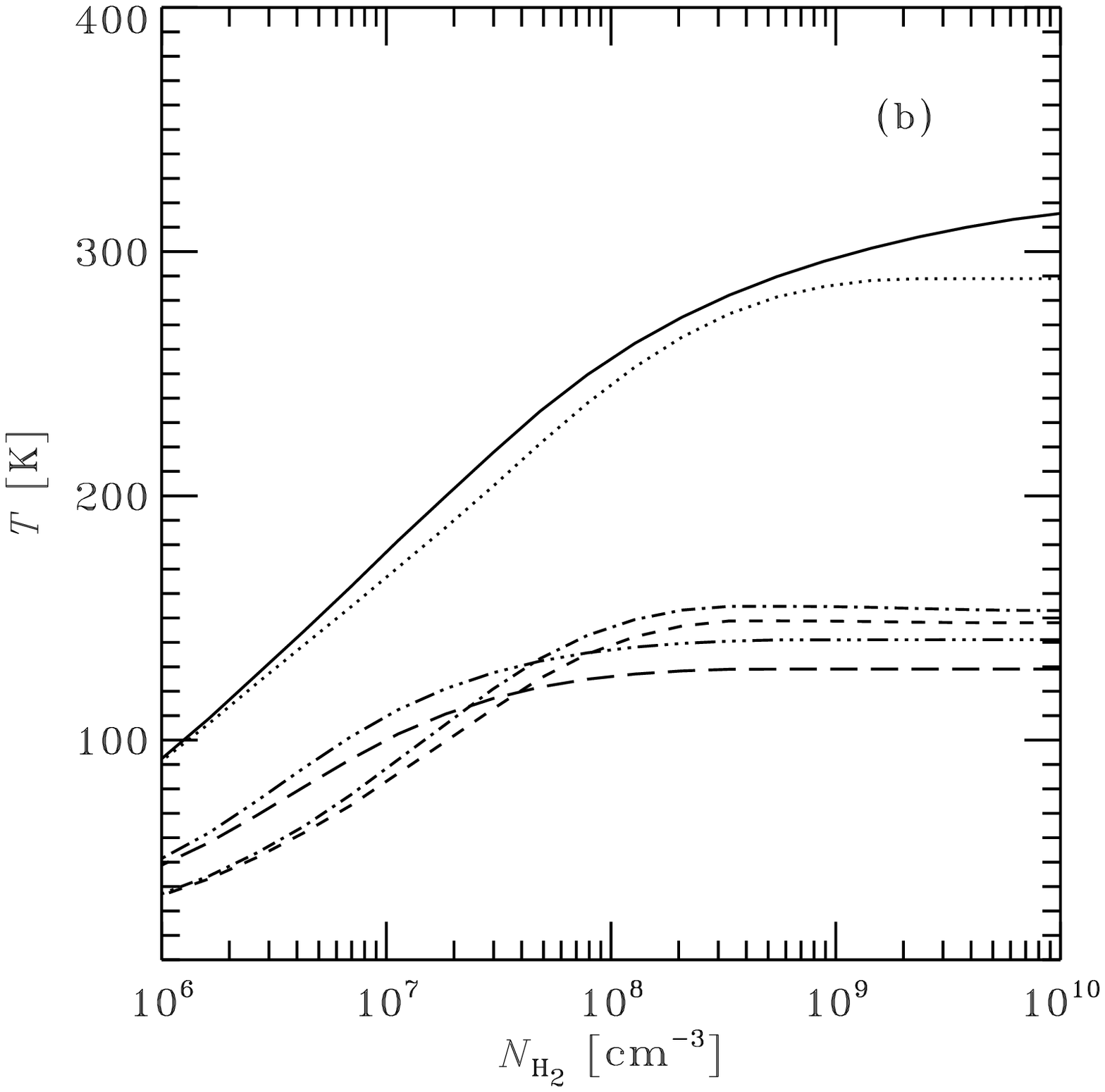}
\end{center}
%\centerline{\epsfig{file=Tm_a_distr.ps,width=8.5 cm}
%\epsfig{file=Cool_Nh_distr.ps,width=8.5 cm}}
\caption { (a)  Dependence of the dust temperature on the grain
size for graphite (solid curve), amorphous carbon (dotted curve),
circumstellar silicate (dashed curve),  astronomical silicate
(dot-dashed curve), crystalline ice (triple-dot-dashed curve),
amorphous ice (long dashed curve).
(b) Gas temperature as a function of the hydrogen concentration
for  grains of the same dust  types as in (a)
 with the size distribution described by Eq.~(\ref{eq:distr}).
Other parameters are  $\fw=10^{-3}$, $\fd=10^{-2}$, $H=10^{14}$
cm,  $W=5 \times 10^{-6}$, $\Tst=3000$ K.}
 \label{fig:Tm_a_distr}
\end{figure*}

The analysis of the population fluxes $F_{ul}$ reveals that,
unlike the cold dust -- hot gas mechanism operating on the pumping
cycle $5_{23} \buildrel45\mum\over\longrightarrow 4_{14}
\buildrel100\mum\over\longrightarrow 5_{05}
\buildrel82\mum\over\longrightarrow6_{16}\;$ (\citealt{De81};
BP04; see dot-dashed lines in Fig.~\ref{fig:levels}),
 the main pumping cycle of the two-dust mechanism is
$5_{23}\buildrel5.5\mum\over\longrightarrow 6_{34}^{\rm E}
\buildrel7.6\mum\over\longrightarrow 7_{43}
\buildrel21\mum\over\longrightarrow~6_{16}$ (dashed lines in
Fig.~\ref{fig:levels}). In other words, the presence of the hot
dust extinguishes maser pumping based on the absorption of 45
$\mum$ photons, because the absorption coefficients of silicates
and amorphous carbon near  45 $\mum$ are similar (see
Fig.~\ref{fig:Kabs}).

Since the absorption coefficient of the amorphous carbon at 5.5
$\mum$ is much larger than that of  silicates, the $5_{23}
\buildrel5.5\mum\over\longrightarrow6_{34}^{\rm E}$ transition is
dominated by the hot dust. On the other hand, there is a peak in
the absorption coefficient of the silicate dust near 20 $\mum$
(see Fig.~\ref{fig:Kabs}) and, therefore, the
$7_{43}\buildrel21\mum\over\longrightarrow6_{16}$ transition is
dominated by the cold dust. Thus,  level $6_{34}^{\rm E}$ is
populated by absorption of the 5.5 $\mum$ radiation from the hot
dust, while the heat sink is realized by absorbing the 21 $\mum$
photons by the cold dust. Moreover, since the Einstein coefficient
for the $6_{34}^{\rm E} \rightarrow 7_{43}$ transition is large
$A=2.4$ $\rm s^{-1}$,
 water molecules are radiatively de-excited from $6_{34}^{\rm E}$ to level $7_{43}$,
closing the cycle of $6_{16} \rightarrow 5_{23}$ maser pumping.

When cold ice is combined with any other hot dust  the maser
effect is extinguished at large $H$ because of the cycle $6_{16}
\buildrel30\mum\over\longrightarrow 7_{25}
\buildrel75\mum\over\longrightarrow 6_{34}
\buildrel49\mum\over\longrightarrow 5_{23}$ (dotted lines in
Fig.~\ref{fig:levels}). This cycle arises due to the gap in the
ice absorption coefficient near 30 $\mum$ (see
Fig.~\ref{fig:Kabs}). The molecules are excited from $6_{16}$ to
$7_{25}$ by absorption of 30 $\mum$ photons from the hot dust,
which dominates this transition. De-excitation from $7_{25}$ to
$5_{23}$ is realized through
$7_{25}\buildrel75\mum\over\longrightarrow 6_{34}$ and
$6_{34}\buildrel49\mum\over\longrightarrow 5_{23}$ transitions
dominated by the cold ice (since its absorption coefficient is
larger than that of silicates or amorphous carbon at these
wavelengths) and inversion disappears.

Fig.~\ref{fig:W_maser} shows the dependence of the maser power on
the dilution factor for the dust combinations (3)--(8) considered
in  Figs.~\ref{fig:f_hc_maser}, \ref{fig:H_maser}. One can see
that the amorphous carbon appears to be more effective than
graphite in producing the inversion of the maser levels, while the
astronomical silicate is more effective than the circumstellar
silicate. It happens  because the absorption coefficient at $21$
$\mum$ ($7_{43} \longrightarrow 6_{16}$ transition responsible for
the heat sink in the maser pumping cycle) of the amorphous carbon
is smaller than that of graphite, while the absorption coefficient
of astronomical silicate is larger than that of circumstellar
silicate (see Fig.~\ref{fig:Kabs}). The larger the efficiency of
absorption of these photons by the cold dust and the weaker the
emission of these  photons by the  hot dust, the larger the
inversion of the maser level populations.

\subsection{Grain size distribution}

\begin{figure}
\centerline{\epsfig{file=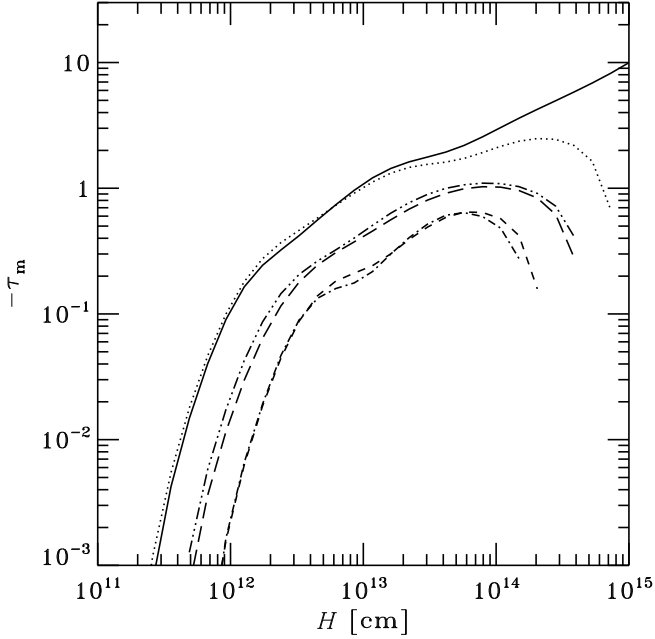,width=8.5 cm}} \caption
{ The optical depth in the $6_{16}-5_{23}$ maser line
 as a function of the half-thickness of the slab
for the same dust types as in Fig.~\ref{fig:Tm_a_distr} and the
same parameters, $\Nh=10^{8}$ cm$^{-3}$. }
\label{fig:H_maser_distr}
\end{figure}

In the previous section we showed that the presence of two types
of dust solves the problem of gas heating as well as provides the
necessary conditions for strong maser pumping. A similar effect
can be achieved if there is a dust grain size distribution. In
Fig.~\ref{fig:Tm_a_distr}a we present the dependence of the dust
temperature  on grain  size. We see that the silicate, graphite
and amorphous carbon temperatures vary a lot, while the ice
temperature is almost constant, because  optical properties of the
ice weakly depend on the grain size (see Fig.~\ref{fig:Kabs}). The
radiative cooling  by water molecules strongly affects the gas
temperature at small $\Nh$ where collisions with dust are less
efficient (see Fig.~\ref{fig:Tm_a_distr}b). At $\Nh \gtrsim
10^{8}$ cm$^{-3}$ the gas temperature approaches the dust
temperature of the smallest grains which dominate the gas heating
(see Eq.~\ref{eq:dist_bal}). At $\Nh \simeq 10^{7}$ cm$^{-3}$ (for
silicates at $10^{8}$ cm$^{-3}$), the gas temperature becomes
smaller than the minimum dust temperature.

\begin{figure}
\centerline{\epsfig{file=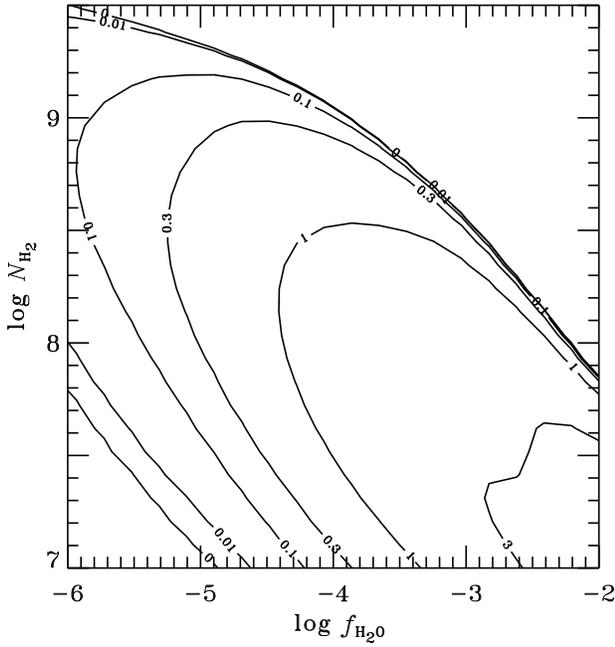,width=8.5 cm}} \caption
{Contour plots of levels of  constant $6_{16} \rightarrow 5_{23}$
maser optical depth, $-\tau_{\rm m}$, for the amorphous carbon
dust   distribution (\ref{eq:distr}). Same parameters as in
Fig.~\ref{fig:Tm_a_distr}.} \label{fig:Nh_ata_distr}
\end{figure}

In Fig.~\ref{fig:H_maser_distr} we present the dependence of the
maser optical depth   as a function of the slab half-thickness.
The maser absorption coefficient is maximal for
$H=10^{12}-10^{13}$ cm and disappears at   $H\lesssim 10^{11}$ cm.
It can be explained as follows. The smaller  $H$, the smaller is
the trapping of photons in the slab and the larger is the
radiative cooling of the gas by water. At small $H$ the gas
temperature thus becomes too low for maser pumping and the maser
disappears. For ice, silicate and amorphous carbon the inversion
exists only for $10^{11}\lesssim H\lesssim 10^{15}$ cm, which
means that here the maser is  pumped due to the de Jong mechanism.
Graphite and carbon dust have the highest temperatures and largest
temperature variations. This causes the maser to operate even in
the deep layers. We see also that the maser is stronger for ice
dust than for the silicate, in spite of the fact that the ice
temperature is smaller than of the silicates (see
Fig.~\ref{fig:Tm_a_distr}a). The reason is that the water cooling
rate is larger in the presence of silicates than in the presence
of ice. It happens because the ice  traps radiation better than
silicates, since $\rm H_2O$ lines are mostly concentrated in the
IR region, where the absorption coefficient of ice is larger than
that of silicates.

The dependence of maser optical depth $\tau_{\rm m}$ on the
hydrogen concentration and water-to-gas mass ratio is shown in
Fig.~\ref{fig:Nh_ata_distr} for the amorphous carbon dust
distribution. The maser strength is large near $\Nh \simeq 10^7
\div 10^8$ cm$^{-3}$ and high water concentration. These results
are similar to that for the case of the two-dust model (see
Fig.~\ref{fig:Nh_ata_contur}), because the cycle of the maser
pumping in these two models is the same.

\subsection{Applications}

\subsubsection{Silicate carbon star V778 Cyg}
\label{sec:v778cyg}

Water masers are observed in some silicate carbon stars, e.g.
V778 Cyg \citep{En94,EL94}. This object also shows silicate
dust emission features in the IR spectra \citep{YD00}.
Recent detailed mapping of the masers with MERLIN revealed
an elongated S-shape structure of about 32 AU   \citep{SS05}.
Such a structure can be interpreted as a warped disk around a
low-mass companion of the carbon star in a binary system observed
almost edge-on.

The  lower limit on the maser brightness temperature is
$T_{\rm B} = 6 \times 10^8$ K. This requires the maser optical
depth of at least $\tau=-\ln [T_{\rm B} / T_{\rm ex}]\approx -15.6$,
where the excitation temperature is  $\sim100$ K.
\citet{SS05} estimated the distance between the masing disk
and the carbon star to be about 80 AU. With the stellar radius
of 2 AU, the dilution factor is about $W= 10^{-4}$.

We assume that the disk contains a mixture of 0.01$\mum$
astronomical silicate and amorphous carbon grains with the total
mass fraction $\fd=0.01$. In the stellar radiation field of a
carbon star (with temperature of about 3000 K), the temperature of
the dust is $T_{\rm as}=235$ K and   $T_{\rm ac}=500$ K,
respectively. Taking $\Nh=10^{8}$ cm$^{-3}$ and  water
fraction $\fw=6\times 10^{-3}$ \citep{JW03}, the resulting gas
temperature is $T=360$ K. From Fig.~\ref{fig:H_maser}, we get
$\taum \approx 8$ for $H=10^{14}$ cm which gives us the absorption
coefficient $\alw^{\rm m}=\taum/H=-8\times 10^{-14}$ cm$^{-1}$.
For the projected disk radius of  $R=16$ AU, the coherent length
in the disk is about $S=1.2 R=3\times 10^{14}$ cm.  Using
Eq.~(\ref{eq:taumax}), we get $\tau=S \; \alw^{\rm m}/\sqrt{\pi} =
-13.5$.  However, for $H=5\times 10^{13}$ cm (e.g. away from the
disk central plane), the inversion is larger by a factor 1.5 (see
Fig.~\ref{fig:W_maser}). This then gives the maximum maser optical
depth of $-20$ which is larger than the observed lower limit. Even
for a much lower water content $\fw=10^{-3}$ we still get
$\tau=-14$. Thus, collisional energy exchange with the dust can
provide the gas heating which is necessary for the masers to
operate.

\subsubsection{Masers from AGB winds}
\label{sec:agb_winds}

Water masers from AGB stars are observed in their
expanding envelopes.
Four low-mass late-type stars (IKTau, U Ori, RT Vir and U Her) have
been mapped by MERLIN  recently \citep{BC03}.
The  data show that maser radiation comes from individual maser clouds
with the apparent size of $2-4$ AU \citep{RY99} and the
filling factor of only $\sim 0.01$. Masers are observed at
 typical distance from the star of  $10 - 70$ AU.
The total maser photon production rate (luminosity)
is about $(1 - 8) \times 10^{42}$ s$^{-1}$. With the total
number of clouds varying between 14 and 286 depending on a source,
one can estimate a single cloud luminosity of less than
$10^{41}$ s$^{-1}$. For a spherical cloud with the radius of
$\sim 2.5$ AU, the maser emissivity $\Phi$ varies
between $\sim 0.2$ and $\sim 0.6$ cm$^{-3}$ s$^{-1}$.

\begin{figure}
\centerline{\epsfig{file=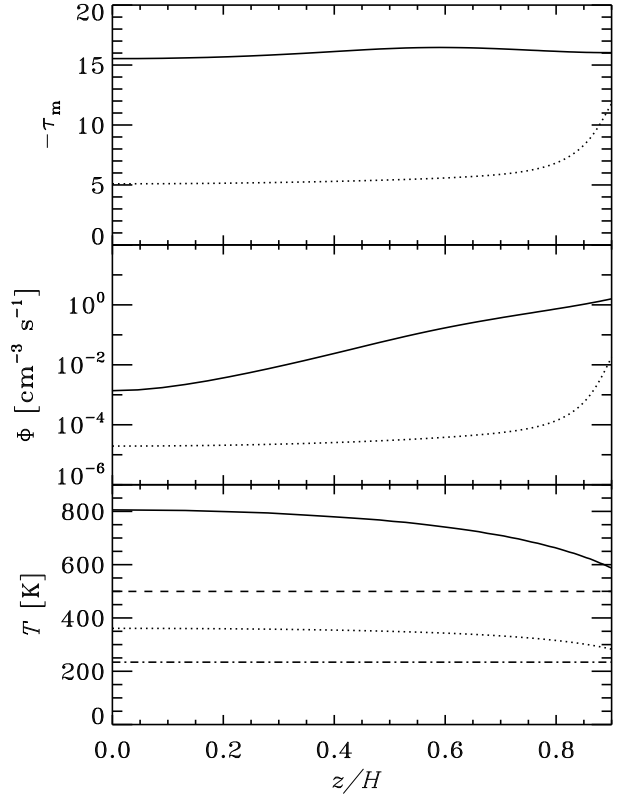,width=8.5cm}} \caption {
Dependence of the maser optical depth, maser photon emissivity
$\Phi$ [photon ${\rm cm^{-3} s^{-1}]}$  and temperatures of the
gas and dust on a height $z$ within a slab ($z=0$ in the slab
center). Solid curves show the case of the drift heating of the
gas given by Eq.~(\ref{eq:heat_dyn})  and dotted curves represent
the case of heating due to thermal motions  (see
Eq.~[\ref{eq:Q+gas}]). Dashed and dot-dashed curves represent the
temperatures of the astronomical silicate and amorphous carbon,
respectively, calculated for $W=10^{-4}$ and $\Tst=3000$ K. Other
parameters are $\Nh=10^{8}$ cm$^{-3}$, $\fw=6 \times 10^{-3}$,
$\fd=10^{-2}$, $H=4 \times 10^{13}$ cm, $a=0.01$ $\mum$, drift
velocity
 $V_{\rm d}=2$ km s$^{-1}$. }
\label{fig:H_str}
\end{figure}

Let us consider a gas - dust cloud of radius $H=4\times 10^{13}$
cm in a radiation field of the star with $\Tst=3000$ K and
$W=10^{-4}$ (i.e. at a distance of about 50 AU for a 1 AU stellar
radius), and other parameters are the same as in section
\ref{sec:v778cyg}. The gas temperature depends on the heating
which can be provided by collisions with dust. We consider
heating  (a) due to the gas thermal motion (see
Eq.[\ref{eq:Q-gas}]) and (b) due to the drift of grains through
the gas (see Eq.[\ref{eq:heat_dyn}]). If only heating due to
thermal motions is considered, the gas temperature is about 360 K
and it lies between temperatures of the cold and hot dust. The
maser optical depth is $\taum\approx -5$ (see dotted curve in
Fig.~\ref{fig:H_str}). From our model we also can compute the
average maser photon emissivity \citep[see e.g.][]{El91}
\begin{eqnarray}
\Phi=g_u A_{ul}\ |\Delta n_{ul}|\ \Nw |S_{ul}| \ K_2^{\rm m},
\label{eq:F_sat}
\end{eqnarray}
which is about $2 \times 10^{-3}$ cm$^{-3}$ s$^{-1}$, i.e.
two orders of magnitudes lower than observed.

When drift heating corresponding to the velocity of only $V_{\rm
d}=2$ km s$^{-1}$ is considered, the gas temperature becomes much
larger than temperatures of the dust. The maser optical depth
increases by a factor of three and the resulting power by two
orders of magnitude reaching $\Phi=0.4$ cm$^{-3}$ s$^{-1}$. This
increase occurs because maser pumping is the most effective when
gas is hotter than the dust (see, for example, BP04).

We can conclude that gas heating by collisions with dust (due to
thermal motion) is not sufficient to produce observed maser
luminosity, while additional drift heating by moving dust is
capable of explaining the masers in AGB winds.

\section{Summary}
\label{sec:concl}

We have considered the maser effect in  a medium consisting of a
mixture of gas (hydrogen and water vapor) and dust of various
types. The gas and dust temperatures and level populations of
water molecule are calculated self-consistently from the system of
population balance equations and thermal balance equations for the
gas and dust in the radiation field of a late-type star.

When dust of different types is present, the gas interacting with
the grains can be heated by one type of dust and is cooled by
another. The gas temperature then takes an intermediate value.
Radiative cooling by water and the presence of hot dust strongly
influence the water molecule energy level populations and
therefore should be taken into account in calculating  the maser
effect. We find that for a  small slab thickness $H$ the inversion
appears because of the de Jong (1973) mechanism, while for large
$H$, the maser can be pumped by radiation from the dust, whose
temperature differs from that of the gas.

The  maser strength depends on the combination of dust types. When
the medium is optically thick to the line radiation, the inversion
of the $6_{16} \rightarrow 5_{23}$ maser level populations appears
only for combinations  of silicates with carbon (or graphite). The
main cycle of maser pumping is
$5_{23}\buildrel5.5\mum\over\longrightarrow 6_{34}^{\rm E}
\buildrel7.6\mum\over\longrightarrow 7_{43}
\buildrel21\mum\over\longrightarrow~6_{16}$. The upward transition
in this cycle is dominated by radiation from the hot dust, while
the heat sink is realized by photon absorption by the  cold dust.
Combinations of water ice with any other dust type produces no
inversion because of the gap in the  ice  absorption coefficient
near 30 $\mum$. Thus,  masers operating on the difference between
the gas and ice  temperatures (\citealt{De81}; BP04) are
extinguished by the hot dust radiation.

Strong masers can also be produced if there is a size distribution
of the dust grains. The maser effect appears due to the de Jong
mechanism for all discussed dust types, if slab half-thickness $H
\lesssim 10^{15}$ cm. We find that the maser disappears at $H
\lesssim 10^{11}$ cm, because the gas temperature becomes too low
due to water cooling. For graphite,  the inversion exists also at
$H \gtrsim 10^{15}$ cm, where the maser pumping  cycle appears to
be the same as in the two-dust maser model.

We show  that  the hot-cold dust model is able to reproduce the
strength of water masers observed from a disk around the companion
of the carbon star in the binary system  V778 Cyg. However, the
masers in the winds of AGB stars require an additional source of
heating, for example due to friction between drifting dust grains and the gas.

\begin{acknowledgements}
This work was supported by the Magnus Ehrnrooth Foundation, the
Finnish Graduate School for Astronomy and Space Physics (NB), and
the Academy of Finland (JP). We are grateful to Dmitrii Nagirner for
the code computing $K$- and $L$-functions,
Ryszard Szczerba for providing dust absorption
coefficients and Seppo Alanko
for useful discussions.
\end{acknowledgements}

%----------------------------------------------------------------------------------------------
\bibliographystyle{apj}
\bibliography{ref}

\begin{thebibliography}{47}
\expandafter\ifx\csname natexlab\endcsname\relax\def\natexlab#1{#1}\fi

\bibitem[{{Babkovskaia} \& {Poutanen}(2004)}]{BP04}
{Babkovskaia}, N. \& {Poutanen}, J. 2004, A\&A, 418, 117 (BP04)

\bibitem[{{Bains} {et~al.}(2003){Bains}, {Cohen}, {Louridas}, {Richards},
  {Rosa-Gonz{\' a}lez}, \& {Yates}}]{BC03}
{Bains}, I., {Cohen}, R.~J., {Louridas}, A., {Richards}, A.~M.~S.,
  {Rosa-Gonz{\' a}lez}, D., \& {Yates}, J.~A. 2003, MNRAS, 342, 8

\bibitem[{{Bertie} {et~al.}(1969){Bertie}, {Labbe}, \& {Whalley}}]{Be69}
{Bertie}, J.~E., {Labbe}, H.~J., \& {Whalley}, E. 1969, J. Chem. Phys., 50,
  4501

\bibitem[{{Bolgova} {et~al.}(1977){Bolgova}, {Strelnitskii}, \&
  {Shmeld}}]{BS77}
{Bolgova}, G.~T., {Strelnitskii}, V.~S., \& {Shmeld}, I.~K. 1977, Soviet
  Astronomy, 21, 468

\bibitem[{{Chandra} {et~al.}(1984{\natexlab{a}}){Chandra}, {Kegel},
  {Varshalovich}, \& {Albrecht}}]{CK84a}
{Chandra}, S., {Kegel}, W.~H., {Varshalovich}, D.~A., \& {Albrecht}, M.~A.
  1984{\natexlab{a}}, A\&A, 140, 295

\bibitem[{{Chandra} {et~al.}(1984{\natexlab{b}}){Chandra}, {Varshalovich}, \&
  {Kegel}}]{CV84b}
{Chandra}, S., {Varshalovich}, D.~A., \& {Kegel}, W.~H. 1984{\natexlab{b}},
  A\&AS, 55, 51

\bibitem[{{Collison} \& {Watson}(1995)}]{CW95}
{Collison}, A.~J. \& {Watson}, W.~D. 1995, ApJ, 452, L103

\bibitem[{{Cooke} \& {Elitzur}(1985)}]{CE85}
{Cooke}, B. \& {Elitzur}, M. 1985, ApJ, 295, 175

\bibitem[{{Danchi} {et~al.}(1994){Danchi}, {Bester}, {Degiacomi}, {Greenhill},
  \& {Townes}}]{DB94}
{Danchi}, W.~C., {Bester}, M., {Degiacomi}, C.~G., {Greenhill}, L.~J., \&
  {Townes}, C.~H. 1994, AJ, 107, 1469

\bibitem[{{David} \& {Pegourie}(1995)}]{DP95}
{David}, P. \& {Pegourie}, B. 1995, A\&A, 293, 833

\bibitem[{{de Jong}(1973)}]{dJ73}
{de Jong}, T. 1973, A\&A, 26, 297

\bibitem[{{de Jong}(1977)}]{dJ77}
---. 1977, A\&A, 55, 137

\bibitem[{{Deguchi}(1977)}]{De77}
{Deguchi}, S. 1977, PASJ, 29, 669

\bibitem[{{Deguchi}(1981)}]{De81}
---. 1981, ApJ, 249, 145

\bibitem[{{Downes} {et~al.}(1981){Downes}, {Genzel}, {Becklin}, \&
  {Wynn-Williams}}]{DG81}
{Downes}, D., {Genzel}, R., {Becklin}, E.~E., \& {Wynn-Williams}, C.~G. 1981,
  ApJ, 244, 869

\bibitem[{{Elitzur}(1991)}]{El91}
{Elitzur}, M. 1991, {Astronomical Masers} (Dordrecht: Kluwer Academic
  Publishers)

\bibitem[{{Engels}(1994)}]{En94}
{Engels}, D. 1994, A\&A, 285, 497

\bibitem[{{Engels} \& {Leinert}(1994)}]{EL94}
{Engels}, D. \& {Leinert}, C. 1994, A\&A, 282, 858

\bibitem[{{Goldreich} \& {Kwan}(1974)}]{GK74}
{Goldreich}, P. \& {Kwan}, J. 1974, ApJ, 191, 93

\bibitem[{{Goldreich} \& {Scoville}(1976)}]{GS76}
{Goldreich}, P. \& {Scoville}, N. 1976, ApJ, 205, 144

\bibitem[{{Goldsmith} \& {Langer}(1978)}]{GL78}
{Goldsmith}, P.~F. \& {Langer}, W.~D. 1978, ApJ, 222, 881

\bibitem[{{Green}(1980)}]{Gr80}
{Green}, S. 1980, ApJS, 42, 103

\bibitem[{{Green} {et~al.}(1993){Green}, {Maluendes}, \& {McLean}}]{GM93}
{Green}, S., {Maluendes}, S., \& {McLean}, A.~D. 1993, ApJS, 85, 181

\bibitem[{{Groenewegen}(1994)}]{Gr94}
{Groenewegen}, M.~A.~T. 1994, A\&A, 290, 531

\bibitem[{{Hartquist} {et~al.}(1980){Hartquist}, {Dalgarno}, \&
  {Oppenheimer}}]{HD80}
{Hartquist}, T.~W., {Dalgarno}, A., \& {Oppenheimer}, M. 1980, ApJ, 236, 182

\bibitem[{{Hollenbach} \& {McKee}(1979)}]{HM79}
{Hollenbach}, D. \& {McKee}, C.~F. 1979, ApJS, 41, 555

\bibitem[{{Hudgins} {et~al.}(1993){Hudgins}, {Sandford}, {Allamandola}, \&
  {Tielens}}]{HS93}
{Hudgins}, D.~M., {Sandford}, S.~A., {Allamandola}, L.~J., \& {Tielens},
  A.~G.~G.~M. 1993, ApJS, 86, 713

\bibitem[{{Humphreys} {et~al.}(2001){Humphreys}, {Yates}, {Gray}, {Field}, \&
  {Bowen}}]{HY01}
{Humphreys}, E.~M.~L., {Yates}, J.~A., {Gray}, M.~D., {Field}, D., \& {Bowen},
  G.~H. 2001, A\&A, 379, 501

\bibitem[{{Jeong} {et~al.}(2003){Jeong}, {Winters}, {Le Bertre}, \&
  {Sedlmayr}}]{JW03}
{Jeong}, K.~S., {Winters}, J.~M., {Le Bertre}, T., \& {Sedlmayr}, E. 2003,
  A\&A, 407, 191

\bibitem[{{Jura}(1996)}]{Ju96}
{Jura}, M. 1996, ApJ, 472, 806

\bibitem[{{Kegel}(1975)}]{Ke75}
{Kegel}, W.~H. 1975, A\&A, 44, 95

\bibitem[{{Laor} \& {Draine}(1993)}]{LD93}
{Laor}, A. \& {Draine}, B.~T. 1993, ApJ, 402, 441

\bibitem[{{Mathis} {et~al.}(1977){Mathis}, {Rumpl}, \& {Nordsieck}}]{MR77}
{Mathis}, J.~S., {Rumpl}, W., \& {Nordsieck}, K.~H. 1977, ApJ, 217, 425

\bibitem[{{Neufeld} {et~al.}(1995){Neufeld}, {Lepp}, \& {Melnick}}]{NL95}
{Neufeld}, D.~A., {Lepp}, S., \& {Melnick}, G.~J. 1995, ApJS, 100, 132

\bibitem[{{Neufeld} \& {Melnick}(1987)}]{NM87}
{Neufeld}, D.~A. \& {Melnick}, G.~J. 1987, ApJ, 322, 266

\bibitem[{{Richards} {et~al.}(1999){Richards}, {Yates}, \& {Cohen}}]{RY99}
{Richards}, A.~M.~S., {Yates}, J.~A., \& {Cohen}, R.~J. 1999, MNRAS, 306, 954

\bibitem[{{Rosen} {et~al.}(1978){Rosen}, {Moran}, {Reid}, {Walker}, {Burke},
  {Johnston}, \& {Spencer}}]{RM78}
{Rosen}, B.~R., {Moran}, J.~M., {Reid}, M.~J., {Walker}, R.~C., {Burke}, B.~F.,
  {Johnston}, K.~J., \& {Spencer}, J.~H. 1978, ApJ, 222, 132

\bibitem[{{Rouleau} \& {Martin}(1991)}]{RM91}
{Rouleau}, F. \& {Martin}, P.~G. 1991, ApJ, 377, 526

\bibitem[{{Shakura} \& {Sunyaev}(1973)}]{SS73}
{Shakura}, N.~I. \& {Sunyaev}, R.~A. 1973, A\&A, 24, 337

\bibitem[{{Sobolev}(1960)}]{So60}
{Sobolev}, V.~V. 1960, {Moving envelopes of stars} (Cambridge: Harvard
  University Press)

\bibitem[{{Strelnitskij}(1977)}]{St77}
{Strelnitskij}, V.~S. 1977, Soviet Astronomy, 21, 381

\bibitem[{{Szczerba} {et~al.}(2005){Szczerba}, {Szymczak}, {Babkovskaia},
  {Poutanen}, {Richards}, \& {Groenewegen}}]{SS05}
{Szczerba}, R., {Szymczak}, M., {Babkovskaia}, N., {Poutanen}, J., {Richards},
  A., \& {Groenewegen}, M. 2005, A\&A, submitted [astro-ph/0504354]

\bibitem[{{Tielens} \& {Hollenbach}(1985)}]{TH85}
{Tielens}, A.~G.~G.~M. \& {Hollenbach}, D. 1985, ApJ, 291, 722

\bibitem[{{Toth}(1991)}]{To91}
{Toth}, R.~A. 1991, J. Opt. Soc. Am. B, 8, 2236

\bibitem[{{Wallin} \& {Watson}(1997)}]{WW97}
{Wallin}, B.~K. \& {Watson}, W.~D. 1997, ApJ, 476, 685

\bibitem[{{Yamamura} {et~al.}(2000){Yamamura}, {Dominik}, {de Jong}, {Waters},
  \& {Molster}}]{YD00}
{Yamamura}, I., {Dominik}, C., {de Jong}, T., {Waters}, L.~B.~F.~M., \&
  {Molster}, F.~J. 2000, A\&A, 363, 629

\bibitem[{{Yates} {et~al.}(1997){Yates}, {Field}, \& {Gray}}]{YF97}
{Yates}, J.~A., {Field}, D., \& {Gray}, M.~D. 1997, MNRAS, 285, 303

\end{thebibliography}

%\clearpage

\end{document}